\documentclass[12pt]{article}
\usepackage{amsmath}
\usepackage{amssymb}
\usepackage{amsthm}
\usepackage{psfrag}
\usepackage{graphicx}

\newcommand{\be}{\begin{equation}}
\newcommand{\ee}{\end{equation}}
\newcommand{\beqa}{\begin{eqnarray}}
\newcommand{\eeqa}{\end{eqnarray}}
\newcommand{\LL}{{\cal L}}

\renewcommand\l{\lambda}
\newcommand\m{\mu}
\newcommand\D{\Delta}
\newcommand\n{\nu}

\newcommand\s{\sigma}

\renewcommand\a{\alpha}
\newcommand\vk{\varkappa}
\newcommand\PP{{\cal P}}
\newcommand\OO{{\cal O}}
\newcommand\Det{{\rm Det\; }}

\def\e{{\rm e}}
\def\d{\partial}
\newcommand{\bseq}{\begin{subequations}}
\newcommand{\eseq}{\end{subequations}}

\newcommand{\di}{\mathrm d}

\begin{document}
\begin{titlepage}
\clearpage

\title{\vspace{-2cm} 
\begin{flushright}
{\normalsize
CERN-PH-TH-2015-287}, 
%\vspace{-0.5cm}
{\normalsize INR-TH-2015-032}\\
\vspace{-0.5cm}
{\normalsize IFT-UAM/CSIC-15-127},
%\vspace{-0.5cm}
{\normalsize FTUAM-15-44},
{\normalsize FR-PHENO-2015-015}
\end{flushright}
\vspace{0.5cm} 
{\bf Renormalization of Ho\v rava Gravity}}

\author{Andrei O.~Barvinsky$^{1,2,3}$, Diego Blas$^{4}$, 
Mario Herrero-Valea$^{5,6}$,\\
Sergey M.~Sibiryakov$^{4,7,8}$, Christian F.~Steinwachs$^9$\\[2mm]
{\small\it $^1$ Theory Department, Lebedev Physics Institute, }%\\[-1mm]
{\small \it  Leninsky Prospect 53, Moscow 119991, Russia}\\
{\small \it $^2$ Department of Physics, Tomsk State University,
  Lenin Ave. 36, Tomsk 634050, Russia}\\
{\small\it $^3$ Pacific Institue for Theoretical Physics,
  Department of Physics and Astronomy, UBC,}\\ [-1mm]
{\small\it
6224 Agricultural Road, Vancouver, BC V6T 1Z1, Canada}\\
{\small\it $^4$ Theoretical Physics Department, CERN 
CH-1211 Geneva 23, Switzerland}\\
{\small\it $^5$ Instituto de F\' isica Te\'orica UAM/CSIC,}\\[-1mm]
{\small\it C/ Nicolas Cabrera, 13-15, 
C.U. Cantoblanco, 28049 Madrid, Spain}\\
{\small\it $^6$ Departamento\! de\! F\'isica\! Te\'orica,\!
Universidad\! Aut\'onoma\! de\! Madrid,\! 20849\! Madrid,\! Spain}\\
{\small\it $^7$ Laboratory of Particle Physics and Cosmology, 
Institute of Physics,} \\
{\small\it
 EPFL, CH-1015 Lausanne, Switzerland}\\
{\small\it $^8$ Institute for Nuclear Research of the
Russian Academy of Sciences,}\\[-1mm]
{\small\it 60th October Anniversary Prospect, 7a, 117312
Moscow, Russia}\\
{\small\it $^9$ Physikalisches Institut,
  Albert-Ludwigs-Universit\"at 
Freiburg,}\\[-1mm]
{\small\it Hermann-Herder-Strasse 3, 79104 Freiburg, Germany}
}
\date{}
\maketitle

\begin{abstract}
We prove perturbative renormalizability of {\em projectable} Ho\v rava
gravity. The key element of the argument is the choice of a gauge which
ensures the correct anisotropic scaling of the propagators and their
uniform falloff at large frequencies and momenta. This guarantees that
the counterterms required to absorb the loop divergences are local and
marginal or relevant with respect to the anisotropic scaling. 
Gauge invariance of the counterterms is achieved by making use of
the background-covariant formalism. We also comment on the difficulties of this approach when addressing the renormalizability of the 
{\em non-projectable} model. 
\end{abstract}

\thispagestyle{empty}
\end{titlepage}

%%%%%%%%%%%%%%%%%%%%%%%%%%%%%%%%%
\section{Introduction}
\label{sec:1}
%%%%%%%%%%%%%%%%%%%%%%%%%%%%%%%%%

The construction of a consistent theory of quantum gravity has remained
one of the major challenges in theoretical physics for many decades.  String
theory provides a fruitful approach to this problem, see e.g. \cite{Polchinski}, at the expense of introducing
a very rich extra structure (and complexity) and it makes sense to question if other directions are possible. In particular, one may wonder whether gravity can
be quantized in the framework of perturbative quantum field theory in 4-dimensions, as other
fundamental forces in Nature. 

At low energies gravity is very well described by
the Einstein--Hilbert action, which is perturbatively non-renormalizable and
therefore does not correspond to an ultraviolet (UV) complete
theory (at least in perturbation theory). It has been known for several decades that a renormalizable
theory is obtained by augmenting the action with quadratic curvature
invariants~\cite{Stelle:1976gc}. For certain regions in the parameter space
the theory is even asymptotically free and hence UV
complete~\cite{Fradkin:1981hx,Avramidi:1985ki}. However, due to the presence of
four time-derivatives of the metric in the Lagrangian, the theory
contains ghosts --- negative-norm states --- and does not admit the
usual interpretation along the lines of unitary quantum
mechanics.\footnote{See \cite{Salvio:2014soa,Einhorn:2014gfa} 
for recent revival of
  this idea.}

An interesting development was proposed by   
P.~Ho\v rava \cite{Horava:2008ih,Horava:2009uw}, 
who pointed out that unitarity can
be preserved at the expense of sacrificing the Lorentz invariance
(LI). In this case one can keep the action to be second order in
time-derivatives, supplementing it only with terms containing higher
spatial derivatives. This allows to construct an action for gravity
which is power-counting renormalizable, i.e. it contains only marginal
and relevant operators with respect to the scaling transformations 
\be
\label{scaling}
t\mapsto b^{-d}\,t~,~~~
x^i\mapsto b^{-1}\, x^i~,~~i=1,\ldots,d\;,
\ee 
where $b$ is an arbitrary scaling parameter and $d$ is the number
of spatial dimensions. Note that time and space scale differently in
(\ref{scaling}). This type of transformations is called anisotropic scaling or
Lifshitz scaling. The metric has zero scaling dimension under\footnote{To be
  precise, this applies to the spatial components of the metric, see
  Sec.~\ref{sec:2} for details.}~(\ref{scaling}),
\[
%\label{hscaling}
\gamma_{ij}\mapsto \gamma_{ij}\;,
\] 
and thus the non-linearities of gravity do not give rise to any irrelevant interactions.

Ho\v rava's proposal generated a surge of papers exploring its low-energy
consistency and phenomenology, see
\cite{Mukohyama:2010xz,Sotiriou:2010wn} for
reviews. This led to the identification of a version of the proposal ---
the so-called healthy non-projectable model~\cite{Blas:2009qj} --- which provides a
consistent theory capable of reproducing the phenomenology of general
relativity (GR) at the distance scales where
the latter has
been tested. It has been also realized that the theory never reduces
to GR exactly: a certain amount of Lorentz invariance violation
persists in the gravity sector at all energy/distance scales
\cite{Blas:2010hb}. 
This
can have interesting implications for cosmological models of dark 
energy~\cite{Blas:2011en}. Conservatively, one
can use astrophysical and cosmological data to constrain the
parameters of the theory
\cite{Blas:2012vn,Yagi:2013qpa,Blas:2014aca}.  
Last but not least, to be phenomenologically viable, this scenario should be
supplemented by a mechanism ensuring Lorentz invariance in the sector
of visible matter where it has been tested with utmost
precision. This represents a serious challenge that
several proposals try to 
address~\cite{GrootNibbelink:2004za,Pujolas:2011sk,Pospelov:2010mp,Bednik:2013nxa}.  

Besides application to gravitation in $4$ dimensions, it was suggested that 
Ho\v rava gravity in $d=2$ can govern the dynamics of 
membranes in M-theory \cite{Horava:2008ih}. Other
uses include the holographic description of non-relativistic
strongly coupled systems, analogous to those occurring in condensed matter 
physics~\cite{Janiszewski:2012nb,Griffin:2012qx}.

Despite the vast literature on Ho\v rava gravity, its renormalizability
has not yet been rigorously proven. Indeed, while in pure scalar and fermionic Lifshitz theories
with non-negative scaling dimensions of the fields\footnote{In
  theories containing fields with negative dimensions the situation is
more subtle~\cite{Fujimori:2015mea}.}
renormalizability is a rather straightforward consequence of
power-counting renormalizability~\cite{Anselmi:2007ri}, this is not the
case for gauge theories. As we are going to explain below, a general local
gauge fixing in Ho\v rava gravity gives rise to certain ``irregular''
contributions in the propagator of the metric, that may spoil the
convergence of the loop integrals (see~\cite{Anselmi:2008bq} for a similar phenomenon in non-relativistic gauge theories). 
As a consequence, a loop diagram that by a scaling argument should be finite can actually diverge and generate a counterterm not expected from the naive power-counting.
Moreover, the
irregular terms in the propagators can potentially lead to
non-local divergences. Hence, the key question is whether there exists a class of
gauges where all propagators are regular.

In this paper we answer this question in the affirmative for the case of
{\it projectable} Ho\v rava gravity. Unfortunately, this version of Ho\v rava gravity does not
reproduce GR at low energies (at least not within weak
coupling)~\cite{Blas:2010hb}. Nevertheless, it presents an interesting example of a
theory sharing many properties of GR, such as a large gauge group of
local spacetime transformations and the presence of gapless
transverse-traceless excitations --- gravitons --- in dimensions $d=3$
and higher. Working in the gauge with regular propagators we demonstrate, with methods along the lines of
relativistic gauge theories,
that projectable Ho\v rava gravity is perturbatively renormalizable in the strict sense. 

In the non-projectable case, we find that there is no gauge fixing which could remove all irregular contributions, though they can be reduced to only a few terms in the  
propagators for the
lapse function ($(00)$-component of the metric). Physically, these terms are a manifestation of the instantaneous interaction present in the theory~\cite{Blas:2010hb,Blas:2011ni}. 
We conclude that the
renormalizability analysis in the non-projectable case requires a
careful treatment of the instantaneous mode.

Previous studies of the quantum properties of Ho\v rava gravity span
several directions. In Ref.~\cite{Orlando:2009en} the
projectable version in $d=3$ is considered with an additional restriction on the
parameters imposed by the condition of detailed balance 
\cite{Horava:2009uw}. This model is connected to 3-dimensional
topologically massive gravity via the stochastic quantization approach and it
is argued that it inherits the renormalizability properties of the
latter. However, the treatment of the gauge invariance of Ho\v rava
gravity in this construction is somewhat obscure. The
works~\cite{Horava:2009if,Ambjorn:2010hu,Sotiriou:2011mu,Benedetti:2014dra} 
explore the
relation between Ho\v rava gravity and causal dynamical
triangulations. In Ref.~\cite{Benedetti:2013pya} a one-loop
renormalization has been performed and the corresponding beta-function were computed in a truncated version
of the $d=2$ projectable model. The truncation, however, explicitly breaks
the gauge invariance of the theory. Finally, in Refs.~\cite{D'Odorico:2014iha} the one-loop counterterms for the
gravitational effective action induced by a scalar field with
Lifshitz scaling (see also
\cite{Giribet:2010th,Nesterov:2010yi,Baggio:2011ha}
for earlier works on this subject) were computed. 
These counterterms were shown to have the same
structure as the terms present in the bare action of Ho\v rava gravity, which suggests that
 if pure Ho\v rava gravity is renormalizable, it
remains so upon inclusion of matter. We will return to this point in the Conclusions (Sec.~\ref{sec:7}).

The paper is organized as follows. In Sec.~\ref{sec:2} we introduce
the projectable version of Ho\v rava gravity in $(3+1)$ and $(2+1)$ dimensions. 
The $d=2$ case
provides the simplest example of gravity with anisotropic scaling,
which we use to illustrate the main ideas of our
approach. In Sec.~\ref{sec:3} we discuss the irregular terms arising in
the propagators for a generic choice of gauge and the associated
problems in the renormalization analysis. In Sec.~\ref{sec:4} we
present a two-parameter family of gauges where the propagators are
free from irregular contributions. Using this class of regular gauges we
evaluate the degree of divergence of a generic diagram in
Sec.~\ref{sec:5} and argue that only local
counterterms that are 
relevant or marginal with respect to the anisotropic scaling are required to
renormalize the theory. By embedding our gauge-fixing procedure into the
background-field formalism, we ensure that the counterterms
preserve gauge invariance, which completes the proof of
renormalizability. In Sec.~\ref{sec:6} we analyze the non-projectable
case and identify irregular contributions that cannot be removed by
a gauge-fixing. We conclude in Sec.~\ref{sec:7}. Some details of the derivations are
relegated to the appendices.

%%%%%%%%%%%%%%%%%%%%%%%%%
\section{Projectable Ho\v rava gravity}
\label{sec:2}
%%%%%%%%%%%%%%%%%%%%%%%%%

Geometrically, Ho\v rava gravity differs from GR by the
introduction of a preferred spacetime foliation by space-like
surfaces. The spacetime metric is represented using the
Arnowitt--Deser--Misner (ADM) decomposition,
\[
%\label{ADM}
\di s^2=N^2\di t^2-\gamma_{ij}(\di x^i+N^i\di t)(\di x^j+N^j\di t)~,~~~~i,j=1,\ldots,d\;.
\]
We aim to construct the theory which is invariant under the subgroup of
diffeomorphisms that preserve the foliation structure (FDiffs). These
consist of time-dependent transformations of the spatial coordinates
and space-independent reparameterizations of time,
\[
%\label{FDiffs}
x^i\mapsto\tilde x^i({\bf x},t)~,~~~~t\mapsto \tilde t(t)\;,
\] 
where $\tilde t(t)$ is a monotonic function.
Under this symmetry the lapse $N$, the shift $N^i$ and the spatial
metric $\gamma_{ij}$ transform in the standard way,
\be
\label{Ftrans}
N\mapsto \tilde N=N\frac{\di t}{\di\tilde t}~,~~~
N^i\mapsto \tilde N^i=\bigg(N^j\frac{\d\tilde x^i}{\d x^j} 
-\frac{\d\tilde x^i}{\d t}\bigg)\frac{\di t}{\di\tilde t}~,~~~
\gamma_{ij}\mapsto\tilde\gamma_{ij}=\gamma_{kl}
\frac{\d x^k}{\d\tilde x^i} \frac{\d x^l}{\d\tilde x^j}\;. 
\ee
We also impose time-reversal invariance, under which $N$ and
$\gamma_{ij}$ are even, whereas the shift $N^i$ is odd.
 
We assign the following scaling dimensions to the fields according to their transformation under the anisotropic scaling\footnote{We assign dimension $-1$
  to the spatial coordinates $x^i$. Accordingly, time has dimension
  $-d$. A field $\Phi$ with dimension $r$ transforms under the scaling
  (\ref{scaling}) as $\Phi\mapsto b^{r}\Phi$.} (\ref{scaling}),
\[
%\label{dimensions}
[N]=[\gamma_{ij}]=0~,~~~~[N^i]=d-1\;.
\] 
The action is constructed from local operators that transform as
scalars
under FDiffs
and have dimension up to $2d$,
\be
\label{genact}
S=\frac{1}{2\vk^2}\int \di t\di ^dx\sqrt{\gamma}N\big(K_{ij}K^{ij}-\l
K^2-{\cal V}\big)\;.
\ee
Here, $\vk^2$, $\l$ are free parameters and the extrinsic curvature of the foliation leaves is given by
\[
%\label{extrcur}
K_{ij}=\frac{\dot\gamma_{ij}-\nabla_iN_j-\nabla_jN_i}{2N}\;.
\]
The trace is defined as $K=\gamma^{ij}K_{ij}$. The dot stands for a time-derivative, indices are raised and lowered by the spatial metric
$\gamma_{ij}$ and the covariant spatial derivatives $\nabla_{i}$ are compatible with
$\gamma_{ij}$. The potential term ${\cal V}$ consists of all allowed combinations of
local invariants of scaling dimension up to $2d$ that are 
made of $\gamma_{ij}$, $N$ and their derivatives with respect to $\nabla_{i}$. In this way one obtains a
Lagrangian consisting of marginal and relevant operators with respect
to the anisotropic scaling which in this sense is power-counting
renormalizable. 

In the {\it non-projectable} Ho\v rava gravity the lapse $N$ is assumed
to be a function of both space and time; we postpone the discussion of
this case until Sec.~\ref{sec:6}. For the time being we focus on the
{\it projectable} model where the lapse is a function of time only,
$N=N(t)$. Then the time-reparameterizations allow to set $N=1$ leaving
the time-dependent spatial diffeomorphisms as the remaining gauge
transformations. 

In $d=3$, upon using the Bianchi identities and integration by parts,
one finds the most general potential~\cite{Sotiriou:2009gy},
\be
\label{Vd3}
\begin{split}
{\cal V}^{d=3}=&\;2\Lambda-\eta R+\m_1R^2+\m_2R_{ij}R^{ij}\\
&+\n_1R^3+\n_2RR_{ij}R^{ij}+\n_3R^i_jR^j_kR^k_i
+\n_4\nabla_iR\nabla^iR+\n_5\nabla_iR_{jk}\nabla^iR^{jk}\;.
\end{split}
\ee   
Here, $R_{ij}$ and $R$ are the Ricci tensor and Ricci scalar
constructed from $\gamma_{ij}$.
In total, the theory contains 11 couplings: $\vk^2$, $\l$, $\Lambda$,
$\eta$, $\m_{1,2}$ and $\n_a$, $a=1,\ldots,5$. The terms in the second
line of (\ref{Vd3}) together with the extrinsic-curvature terms in (\ref{genact})
are marginal under the scaling (\ref{scaling}). They determine the UV
behavior of the theory, in particular its renormalizability
properties. The rest of the terms in (\ref{Vd3}) are relevant
deformations. Among them the cosmological constant $\Lambda$, which has the 
lowest dimension. We will assume that it is tuned to zero in order to admit
 flat Minkowski spacetime as a solution.

Let us study the spectrum of linear perturbations around this
background. We write 
\[
%\label{metrpert}
\gamma_{ij}=\delta_{ij}+h_{ij},
\]
and decompose the perturbations into scalar, vector and
transverse-traceless (TT) tensor parts,
\bseq
\label{metrlin}
\begin{align}
\label{metrlin1}
&N^i=\d_iB+u_i\;,\\
\label{metrlin2}
&h_{ij}=\left(\delta_{ij}-\frac{\d_i\d_j}{\Delta}\right)\psi+
\frac{\d_i\d_j}{\Delta}E+\d_iv_j+\d_jv_i+\zeta_{ij}
\end{align}
\eseq
with
\[
%\label{TTs}
\d_iu_i=\d_iv_i=\d_i\zeta_{ij}=\zeta_{ii}=0\;.
\]  
Here $\D$ is the flat-space Laplacian.
The quadratic action reads,
\be
\label{S2d3}
\begin{split}
S^{d=3}_2=\frac{1}{2\vk^2}\int \di t \di ^3x\bigg[
&\frac{\dot\zeta_{ij}^2}{4}+\frac{\eta}{4}\zeta_{ij}\D\zeta_{ij}
-\frac{\m_2}{4}\zeta_{ij}\D^2\zeta_{ij}
+\frac{\n_5}{4}\zeta_{ij}\D^3\zeta_{ij}
-\frac{1}{2}(\dot v_i-u_i)\D(\dot v_i-u_i)\\
&+\frac{\dot\psi^2}{2}+\frac{1}{4}(\dot E-2\D B)^2
-\frac{\l}{4}(2\dot\psi+\dot E-2\D B)^2\\
&-\frac{\eta}{2}\psi\D\psi
-\bigg(4\m_1+\frac{3\m_2}{2}\bigg)\psi\D^2\psi
+\bigg(4\n_4+\frac{3\n_5}{2}\bigg)\psi\D^3\psi\bigg]\;.
\end{split}
\ee
In order to identify the physical degrees of freedom we perform the variation
with respect to $u_i$ and $B$ and set them to zero afterwards by the gauge
choice. We obtain the equations,
\be
\label{uBeqd3}
\D\dot v_i=0~,~~~~\D\bigg(\dot E-\frac{2\l}{1-\l}\dot\psi\bigg)=0\;.
\ee
The first one implies that the vector sector does not contain any
propagating modes. From the second equation in (\ref{uBeqd3}) we
express $\dot E$ and substitute it back into (\ref{S2d3}) which yields
the action for the propagating degrees of freedom,
\be
\label{S2d3phys}
\begin{split}
S^{d=3}_2=\frac{1}{2\vk^2}\int\!\! \di t \di ^3x\bigg[
&\frac{\dot\zeta_{ij}^2}{4}+\frac{\eta}{4}\zeta_{ij}\D\zeta_{ij}
-\frac{\m_2}{4}\zeta_{ij}\D^2\zeta_{ij}
+\frac{\n_5}{4}\zeta_{ij}\D^3\zeta_{ij}
\\
&+\frac{1-3\l}{2(1-\l)}\dot\psi^2-\frac{\eta}{2}\psi\D\psi
-\bigg(4\m_1+\frac{3\m_2}{2}\bigg)\psi\D^2\psi
+\bigg(4\n_4+\frac{3\n_5}{2}\bigg)\psi\D^3\psi\bigg]\,.
\end{split}
\ee
In addition to the TT mode $\zeta_{ij}$, the theory propagates a ``scalar
graviton'' $\psi$. Both modes have positive-definite kinetic terms
provided $\vk^2>0$ and $\l$ is either smaller than $1/3$ or larger
than $1$. The dispersion relations of the two modes are respectively,
\bseq
\label{dispreld3}
\begin{align}
\label{dispreld3tt}
&\omega_{tt}^2=\eta k^2+\m_2k^4+\n_5k^6\;,\\
\label{dispreld3s}
&\omega_{s}^2=\frac{1-\l}{1-3\l}\big(-\eta k^2+(8\m_1+3\m_2)k^4
+(8\n_4+3\n_5)k^6\big)\;.
\end{align}
\eseq
This immediately raises a problem: the term proportional to $k^2$ in the dispersion
relation cannot be positive for both modes simultaneously. Thus,
non-zero $\eta$ leads to an instability of the Minkowski background
with respect to inhomogeneous perturbations. For positive values of the
parameters $\m_{1,2}$ and $\n_{4,5}$ the instability is cut off at
large spatial momenta and therefore does not affect the UV properties
of the theory. Moreover, we can stabilize the Minkowski spacetime by
simply tuning $\eta$ to zero. However, in that case the dispersion
relations of the TT mode and scalar gravitons are quadratic,
$\omega\propto k^2$, down to zero momentum, which prevents from
recovering GR at low energies\footnote{One could try to keep $\eta$
  finite and positive and suppress the instability associated to the
  scalar graviton by tuning $\l$ close to $1$. However, in this limit
  the theory becomes strongly coupled and the perturbative treatment
  breaks down~\cite{Blas:2009yd,Blas:2010hb}.}.

The situation is much simpler for $d=2$. In this case the potential
includes only two terms,
\be
\label{Vd2}
{\cal V}^{d=2}=2\Lambda+\m R^2\;.
\ee
The linear in $R$ term is absent because the combination
$\sqrt{\gamma}R$ is a total derivative in 2-dimensions. Also the
Ricci tensor $R_{ij}$ reduces to the scalar curvature, so the
invariant $R_{ij}R^{ij}$ is proportional to $R^2$. Setting the
cosmological constant $\Lambda$ to zero, we obtain a model with 3
marginal couplings: $\vk^2$, $\l$ and $\m$.

The spectrum of this model is derived along the same lines as
for the $d=3$ case. Expanding around flat spacetime and performing the
decomposition (\ref{metrlin}) -- where now the TT-component
$\zeta_{ij}$ is absent -- we obtain the quadratic action,
\be
\label{S2d2}
S^{d=2}_2=\frac{1}{2\vk^2}\int \di t\di ^2x 
\bigg[-\frac{1}{2}(\dot v_i-u_i)\D(\dot v_i-u_i)
+\frac{\dot\psi^2}{4}+\frac{1}{4}(\dot E-2\D B)^2
-\frac{\l}{4}(\dot\psi+\dot E-2\D B)^2-\m \psi\D^2\psi\bigg]\;.
\ee
We observe that the action for the vector perturbations has exactly
the same structure as in $d=3$, implying that there are no propagating
modes in this sector. In the scalar sector we eliminate $E$ using the
equation obtained upon variation with respect to $B$ and set $B=0$
afterwards.
This yields,
\[
%\label{S2d2phys}
S^{d=2}_2=\frac{1}{2\vk^2}\int
\di t\di ^2x\bigg[\frac{1-2\l}{4(1-\l)}\dot\psi^2
-\m \psi\D^2\psi\bigg]\;.
\]
Unlike GR, which in $(2+1)$ dimensions does not possess any local
degrees of freedom, Ho\v rava gravity propagates a dynamical scalar
mode. The latter has the dispersion relation,
\[
%\label{dispreld2}
\omega^2_s=4\mu \frac{1-\l}{1-2\l}\,k^4\;.
\]
It is well-behaved (i.e. has positive kinetic term and is stable)
if\footnote{We take $\vk^2>0$ to make contact with higher dimensions
  where this condition is required for positivity of the TT mode kinetic energy.}
$\vk^2>0$, $\m>0$ and $\l<1/2$ or $\l>1$.
We make extensive use of the $d=2$ model in what follows.

In order to analyze the renormalizability properties of the theory, from now on we transform to
``Euclidean'' time by the Wick rotation 
\[
t\mapsto\tau=it, \quad \quad \quad N^j\mapsto N_E^j=-i N^j.
\] 
In the following we will omit the subscript ``$E$'' on the Euclidean shift.
The corresponding action differs from
(\ref{genact}) only by the sign of the potential term. 
At the quadratic level this
amounts to flipping the signs of the terms containing $\m_{1,2}$, 
$\n_{4,5}$ in (\ref{S2d3}) and of  
the $\m$-term in (\ref{S2d2}).   

%%%%%%%%%%%%%%%%%%%%%%%%%%%%%%%%%
\section{Local gauge fixing and irregular terms}
\label{sec:3}
%%%%%%%%%%%%%%%%%%%%%%%%%%%%%%%%%

In this section we focus on the theory in $d=2$.
In order to quantize the theory we need to fix the gauge. Finding a suitable
gauge turns out to be non-trivial, as we demonstrate
below. The technical part of the following analysis is straightforward. Upon
adding a gauge-fixing term to the quadratic action (\ref{S2d2}) and
transforming to momentum space, we invert the kinetic matrices
for the scalar and vector perturbations. The propagators of the shift
and the spatial metric are then reconstructed from these helicity components using Eqs.~(\ref{metrlin}),
\bseq
\label{propfields}
\begin{align}
\label{NN}
\langle N^i(p)N^j(-p) \rangle=&\langle u_i u_j\rangle+
 k_i k_j\langle
BB\rangle\;,\\
\langle N^i(p)h_{jk}(-p)\rangle=&
-ik_j\langle u_iv_k\rangle 
-ik_k\langle u_iv_j\rangle 
+ik_i\big(\delta_{jk}-\hat k_j\hat k_k\big)\langle B\psi\rangle
+ik_i\hat k_j\hat k_k\langle BE\rangle\;.
\label{Nh}\\
\langle h_{ij}(p)h_{kl}(-p)\rangle=&
k_ik_k\langle v_j v_l \rangle+
k_jk_k\langle v_i v_l \rangle+
k_ik_l\langle v_j v_k \rangle+
k_jk_l\langle v_i v_k \rangle\notag\\
&+\big(\delta_{ij}-\hat k_i\hat k_j\big)
\big(\delta_{kl}-\hat k_k\hat k_l\big)\langle\psi\psi\rangle
+\big(\delta_{ij}-\hat k_i\hat k_j\big)\hat k_k\hat k_l
\langle\psi E\rangle\notag\\
&+\hat k_i\hat k_j\big(\delta_{kl}-\hat k_k\hat k_l\big)
\langle E\psi\rangle
+\hat k_i\hat k_j\hat k_k\hat k_l\langle EE\rangle\;.
\label{hh}
\end{align}
\eseq
Here we  introduced the notations 
\[
p\equiv(\omega, {\bf k}), \quad \quad \quad  
\hat k_i\equiv k_i/k.
\]
We postpone the discussion of the Faddeev--Popov ghosts coming from the gauge fixing to Sec.~\ref{Sec:FP}.

Let us illustrate the  type of problems connected to the gauge fixing procedure by 
considering as a first trial the gauge 
\be
\label{simple}
N^i=0\;.
\ee
It can be implemented by adding the term 
\[
%\label{Lgfsimple}
{\cal L}_{gf}=\frac{\sigma}{2\vk^2}(N^i)^2
\]
to the Lagrangian and taking the limit $\sigma\to\infty$. 
Alternatively, one can simply set $u_i=B=0$ in (\ref{S2d2}). The
kinetic matrix for the remaining variables is now invertible yielding
the propagators,
\bseq
\label{propssimple}
\begin{align}
\label{prop1simple}
&\langle v_i(p)v_j(-p)\rangle=\frac{2\vk^2}{\omega^2 k^2}
\big(\delta_{ij}-\hat k_i\hat k_j\big)\;,\\
\label{prop2simple}
&\langle
\psi(p)\psi(-p)\rangle=\frac{4\vk^2(1-\l)}{1-2\l}\,\PP_s(p)\;,\\
\label{prop3simple}
&\langle\psi(p)E(-p)\rangle=\frac{4\vk^2\l}{1-2\l}\,\PP_s(p)\;,\\
\label{prop4simple}
&\langle E(p)E(-p)\rangle=\frac{4\vk^2\l^2}{(1-\l)(1-2\l)}\,\PP_s(p)
+\frac{4\vk^2}{(1-\l)\omega^2}\;,
\end{align}
\eseq
where
\be
\label{Ps}
\PP_s(p)=\bigg[\omega^2+4\m\frac{1-\l}{1-2\l}k^4\bigg]^{-1}
\ee
has the pole corresponding to the physical mode\footnote{Recall that we are working in the Euclidean time, so
  the sign of the $\m$-term in (\ref{S2d2}) must be flipped to~``+''.}. 
Note the presence of the transverse
projector in (\ref{prop1simple}) which is implied by the transversality of
$v_i$.
Substituting these expressions into (\ref{hh}) we obtain,
\be
\label{hhsimple}
\begin{split}
\langle
h_{ij}(p)h_{kl}(-p)\rangle=&\frac{4\vk^2(1-\l)}{1-2\l}\delta_{ij}\delta_{kl}
\,\PP_s(p)
+\big(\delta_{ik}\delta_{jl}+\delta_{il}\delta_{jk}
-2\delta_{ij}\delta_{kl}\big)\frac{2\vk^2}{\omega^2}\\
&+16\vk^2\m\bigg[\frac{1-\l}{1-2\l}\big(\delta_{ij}k_k k_l
+k_ik_j\delta_{kl}\big)k^2-k_ik_jk_kk_l\bigg]\frac{\PP_s(p)}{\omega^2}\;.
\end{split}
\ee
In deriving this expression we used the dimensional dependent identity (\ref{identd2})
that can be found in Appendix~\ref{app:A}.

We observe that besides the first contribution proportional to
$\PP_s(p)$, which uniformly decreases whenever $\omega$ or $k$ 
go to infinity, the propagator (\ref{hhsimple})
contains terms of the form $1/\omega^2$ and $O(k^4)\PP_s/\omega^2$
that do not fall off with the spatial momentum. The latter terms are
dangerous as they lead to non-local singularities of the propagator in
position space. For example, the Fourier transform of
the second term in (\ref{hhsimple}) has the form,
\[
%\label{badsimple}
\langle
h_{ij}(\tau,{\bf x})h_{kl}(0)\rangle\ni
-\vk^2\big(\delta_{ik}\delta_{jl}+\delta_{il}\delta_{jk}
-2\delta_{ij}\delta_{kl}\big)\, |\tau|\,\delta^{(2)}({\bf x})\;,
\]
where $\delta^{(2)}$ is the $\delta$-function. This is singular at
${\bf x}=0$ for {\em all} times $\tau$.\footnote{On the contrary, the
  Fourier transform of $\PP_s$ is singular only at $\tau={\bf
    x}=0$. This is explicitly verified using the representation 
\[
%\label{regrepr}
\begin{split}
\int \frac{\di \omega \di ^2k}{(2\pi)^3}\;
\frac{\e^{-i\omega\tau+i{\bf kx}}}{\omega^2+A^2k^4}
&=-\frac{1}{\D}\int\frac{\di ^2k}{(2\pi)^2A}
\e^{-Ak^2|\tau|+i{\bf kx}}\\
&=
\frac{-1}{16\pi^2 A^2|\tau|}\int \di ^2y
\log|{\bf x}-{\bf y}|\exp\bigg[-\frac{y^2}{4A|\tau|}\bigg]\;.
\end{split}
\]
The r.h.s. is smooth together with all its derivatives whenever
$\tau$ or ${\bf x}$ are non-zero.} 
In the perturbative expansion
such contributions will give rise to overlapping singularities that
are non-local in time. In the more familiar momentum-space
representation they correspond to divergences in the loop diagrams
which have a non-polynomial dependence on the external frequency. Unless
these divergences cancel order by order of perturbation theory, they jeopardize the renormalizability by requiring the introduction of counterterms with non-local time dependence\footnote{A hint towards such cancellation comes from considering one-graviton exchange between two external sources. One can check that the irregular contributions drop off from this amplitude if the sources are conserved, as required by FDiff-invariance. This argument is not directly applicable at higher orders of perturbation theory where the sources coupled to the metric are not conserved due to the non-linearities of the theory.}.  Clearly,
even if the cancellation of non-local divergences does take place, it will be increasingly hard to keep track of it at higher loop orders. Thus, we
conclude that the gauge (\ref{simple}) is not suitable for the
analysis of renormalizability. 
 
The gauge (\ref{simple}) is rather special and one might think that the
non-local singularities in the propagators can be avoided once we
allow for a more general gauge-fixing condition. Let us now show that
this is not the case as long as one restricts to local
gauge-fixing terms. The most general term of this class has the form,
\be
\label{local}
\LL_{gf}=\frac{\sigma}{2\vk^2}\,F^i{\cal O}_{ij}F^j\;,
\ee
where $F^i$ is a linear combination of the fields $N^i$, $h_{ij}$
and their derivatives which transforms as a vector under spatial
rotations, while ${\cal O}_{ij}$ is an invertible local operator. In order not to spoil the
scaling properties of the action, the gauge-fixing term should not
introduce any dimensionful couplings with respect to the scaling \eqref{scaling}. This implies that the total
dimension of $\LL_{gf}$ must be $4$, whereas all terms in $F^i$ and
${\cal O}_{ij}$ must scale in the same way. 
A local operator ${\cal O}_{ij}$
can contain only the identity and a finite number of
derivatives, and therefore its scaling dimension is non-negative. This
implies that the dimension of $F^i$ must be less or equal to
$2$. This excludes that $F^i$ can contain time derivatives of the shift, since such terms would already have at least a scaling dimension of $3$. The time derivative of $h_{ij}$ also
cannot appear in $F^i$ because to obtain from it an
object with a single index one must introduce an additional spatial
derivative, which again raises the dimension up to $3$. Finally,
$F^i$ cannot contain $N^i$ and the spatial
derivatives of the metric $h_{ij}$ simultaneously, as otherwise it
would explicitly break the time-reversal invariance\footnote{Note that
  in $d=3$ such combination is further forbidden by the mismatch between the scaling
  dimensions of the shift $N^i$ and of the spatial derivatives of
  $h_{ij}$.}. Thus we arrive at two possibilities:
\be
\label{chilocal}
F^i=N^i~~~~\text{or}~~~
F^i=\d_jh_{ij}+\sigma'\d_i h\;,
\ee
where $\sigma'$ is an arbitrary coefficient and $h$ is the trace of
$h_{ij}$. 
Both these combinations have dimension $1$, so the 
corresponding operator ${\cal O}_{ij}$ must be of dimension $2$.
Hence, it has the form,
\[
%\label{Olocal}
\OO_{ij}=-\delta_{ij}\D-\sigma''\d_i\d_j\;.
\]

Without delving into the study of the full propagators, let us focus
on the transverse component $u_i$ of the shift. For the first choice of the gauge-fixing function in
(\ref{chilocal}) a straightforward
calculation yields,
\[
%\label{ulocal1}
\langle u_i(p)u_j(-p)\rangle=\frac{\vk^2}{\sigma k^2}
\big(\delta_{ij}-\hat k_i\hat k_j\big)
\]
whereas for the second choice one obtains,
\[
%\label{ulocal2}
\langle u_i(p)u_j(-p)\rangle=\bigg[\frac{2\vk^2}{k^2}
+\frac{\vk^2\omega^2}{\sigma k^6}\bigg]
\big(\delta_{ij}-\hat k_i\hat k_j\big)\;.
\]
In both cases the propagator contains contributions independent of the
frequency and behaving as $1/k^2$. This, in turn, leads to a non-local
singularity of the $\langle N^iN^j\rangle$ propagator in the
position space proportional to 
\[
%\label{badlocal}
\delta^{(1)}(\tau)\frac{1}{4\pi} \log|{\bf x}|\;.
\]
In perturbation theory this will produce spurious divergences that are
non-local in space. Therefore, none of the local gauges (\ref{chilocal})
is appropriate for our purposes.
 
%%%%%%%%%%%%%%%%%%%%%%%%%%%%%%%%%
\section{Regular gauges}
\label{sec:4}
%%%%%%%%%%%%%%%%%%%%%%%%%%%%%%%%%

Let us introduce some terminology: 
consider two fields $\Phi_1$, $\Phi_2$ that have scaling dimensions 
$r_1$, $r_2$ under (\ref{scaling}). 
Following \cite{Anselmi:2008bq} we
will denote the propagator $\langle\Phi_1\Phi_2\rangle$
regular if it is given by the sum of terms
of the form,
\bseq
\label{reg}
\be
\label{regform}
\frac{P(\omega,{\bf k})}{D(\omega,{\bf k})}\;,
\ee  
where $D$ is a product of monomials,
\be
\label{regden}
D=\prod_{m=1}^M(A_m\omega^2+B_m k^{2d}+\ldots)~,~~~A_m,B_m>0\;,
\ee
\eseq
and $P(\omega,{\bf k})$ is a polynomial of scaling degree\footnote{The
  scaling degree of a polynomial is defined as the maximal scaling
  dimension of its terms.} less or
equal $r_1+r_2+2(M-1)d$.
We emphasize that all constants $A_m,B_m$ in
(\ref{regden}) must be strictly positive. Ellipsis stands for terms with lower
scaling dimensions that generically arise in theories with relevant
operators in the action. 
The reader will easily convince herself that a regular propagator has only local singularities at $\tau={\bf
  x}=0$ in position space. Due to the restriction on the degree of the numerator it
scales at short distances and time-intervals as 
\[
%\label{propscaling}
\langle\Phi_1(b^{-d}\tau,b^{-1}{\bf x})\,\Phi_2(0)\rangle
=b^{r_1+r_2}\langle\Phi_1(\tau,{\bf x})\,\Phi_2(0)\rangle\;.
\]

The results of the previous section show that in order to
obtain regular propagators in Ho\v rava gravity we need to go beyond
local gauge-fixing terms.

%%%%%%%%%%%%%%%%%%%%%%%%%%%%%%
\subsection{Theory in 2 spatial dimensions}
\label{sec:4a}
%%%%%%%%%%%%%%%%%%%%%%%%%%%%%%

As a starting point and for guidance in the treatment of non-relativistic theories, let us first review the structure of covariant gauges in relativistic theories. In GR and its higher-derivative extensions the
corresponding gauge-fixing Lagrangians can be adjusted such as to cancel the
terms mixing different metrics components in the quadratic action. Such a gauge fixing renders the tensor structure of 
the propagators diagonal and greatly simplifies the
actual computations. The spatial diffeomorphisms are fixed in the
covariant gauges by the conditions $F^{i}=0$. In terms of the ADM variables they 
have the general form,
\be
\label{chicovar}
F^i=\dot N^i-\d_j h_{ij}-C\d_i(2\phi+h)\;,
\ee 
where $\phi$ is the perturbation of the lapse and $C$ is a numerical
constant. Such a form is not appropriate for Ho\v rava gravity, because $\dot
N^i$ and spatial derivatives of $h_{ij}$ have different scaling
dimensions. The discrepancy is easily compensated by
introducing two more spatial derivatives acting on the spatial
metric. Thus, for $d=2$ Ho\v rava gravity we consider the gauge fixing
function of the general form,
\be
\label{chinice}
F^i=\dot N^i -C_1\D\d_j h_{ij}-C_2\D\d_ih-C_3\d_i\d_j\d_k h_{jk}\;. 
\ee  
The coefficients $C_{1,2,3}$ will be chosen shortly to simplify the
quadratic action. Note the presence of the last term in (\ref{chinice})
which does not have an analog in the relativistic case \eqref{chicovar}.

Importantly, with the gauge-fixing function
(\ref{chinice}) the operator $\OO_{ij}$ in the gauge-fixing Lagrangian
(\ref{local}) must have dimension $-2$ and thus is necessarily
non-local. We take,
\be
\label{Onice}
\OO_{ij}=-\big[\delta_{ij}\D+\xi\d_i\d_j\big]^{-1}
=-\frac{\delta_{ij}}{\D}+\frac{\xi}{(1+\xi)}\frac{\d_i\d_j}{\D^2}\;.
\ee
Though unusual, the non-locality of the gauge-fixing
Lagrangian does not introduce any problems in the perturbative
expansion around flat spacetime, as it appears only in the quadratic
action\footnote{It will require a careful treatment, 
however, when we generalize our analysis to 
    the background-field formalism in 
  Sec.~\ref{sec:5b}.}. The only important property at this point is the invertibility of ${\cal O}_{ij}$.

We now choose the coefficients in the linear combination
(\ref{chinice}) in such a way that the contributions coming from
$\LL_{gf}$ cancel the terms mixing $N^i$ and $h_{ij}$ in the quadratic
Lagrangian. The combination with the required properties is ,
\be
\label{chisigxi}
\begin{split}
F^i&=\dot N^i+\frac{1}{2\sigma}\OO_{ij}^{-1}\d_k h_{jk}
-\frac{\l}{2\sigma} \OO_{ij}^{-1}\d_j h\\
&=\dot N^i-\frac{1}{2\s} \D\d_k h_{ik}
+\frac{\l(1+\xi)}{2\s}\D\d_ih-\frac{\xi}{2\s}\d_i\d_j\d_kh_{jk}\;.
\end{split}
\ee  
In this way we arrive at a two-parameter family of gauges depending on
$\s$ and $\xi$. It is instructive to write down the total quadratic
Lagrangian in this $\s\xi$-gauge,
\be
\label{Lnice}
\begin{split}
{\cal L}^{d=2}_2+{\cal L}_{gf}=\frac{1}{2\vk^2}\bigg[&\frac{\dot h_{ij}^2}{4}
-\frac{\l\dot h^2}{4}
-\frac{1}{4\s}\d_j h_{ij}\D\d_kh_{ik}
+\bigg(\m+\frac{\xi}{4\s}\bigg) (\d_i\d_jh_{ij})^2
\\
&-\bigg(2\m+\frac{\l(1+\xi)}{2\s}\bigg)\D h \d_i\d_j h_{ij}
+\bigg(\m+\frac{\l^2(1+\xi)}{4\s}\bigg)(\D h)^2\\
&-\s \dot N^i\big[\delta_{ij}\D+\xi\d_i\d_j\big]^{-1}\dot N^j
+\frac{(\d_iN^j)^2}{2}
+\bigg(\frac{1}{2}-\l\bigg)(\d_i N^i)^2\bigg]\;.
\end{split}
\ee
Note that the non-locality persists only in the term involving
time-derivatives of the shift. 

Inserting again the helicity decomposition (\ref{metrlin}) into the
above Lagrangian, inverting the operators that appear in the resulting
quadratic forms, and combining all contributions in 
(\ref{propfields}) we obtain the propagators,
\bseq
\label{propsnice}
\begin{align}
\label{NNnice}
\langle N^i(p)N^j(-p) \rangle=&\frac{\vk^2}{\s}
(k^2\delta_{ij}-k_ik_j)\,\PP_1(p)
+\frac{\vk^2(1+\xi)}{\s}k_ik_j\,\PP_2(p)\;,\\
\langle h_{ij}(p)h_{kl}(-p)\rangle=
&4\vk^2\delta_{ij}\delta_{kl}\bigg[\frac{1-\l}{1-2\l}
\PP_s(p)-\PP_1(p)\bigg]
+2\vk^2 (\delta_{ik}\delta_{jl}+\delta_{il}\delta_{jk})\PP_1(p)\notag\\
+4\vk^2(\delta_{ij}\hat k_k\hat k_l+&\hat k_i\hat k_j\delta_{kl})
\big[\PP_1(p)-\PP_s(p)\big]
+4\vk^2\hat k_i\hat k_j\hat k_k\hat k_l
\bigg[\frac{1-2\l}{1-\l}\PP_s(p)-2\PP_1(p)+\frac{\PP_2(p)}{1-\l}\bigg]\;,
\label{hhnice}
\end{align}
\eseq
whereas $\langle N^i h_{jk}\rangle$ trivially vanishes.
Here $\PP_s$ is given by the expression (\ref{Ps}) and 
\bseq
\label{P12}
\begin{align}
\label{P1}
\PP_1(p)&=\bigg[\omega^2+\frac{k^4}{2\s}\bigg]^{-1}\;,\\
\label{P2}
\PP_2(p)&=\bigg[\omega^2+\frac{(1-\l)(1+\xi)}{\s}k^4\bigg]^{-1}\;.
\end{align}
\eseq
In deriving Eq.~(\ref{hhnice}) we again made use of the identity
(\ref{identd2}).
The above propagators are regular in the sense of \eqref{regform} provided\footnote{For $\l>1$
  the second condition implies $\xi<-1$. In
  this case the operator (\ref{Onice}) in the gauge-fixing term is not
  positive-definite. We are not aware of any problems related
  to this in the perturbation theory. However, it can lead to 
  complications with the non-perturbative definition of the theory (cf. Eq.~\eqref{weighting}).
\label{foot:12}}
\be
\label{condnice}
\s>0~,~~~~(1-\l)(1+\xi)>0\;.
\ee 
Indeed, (\ref{NNnice}) and the first two terms in
(\ref{hhnice}) are obviously regular. 
For the terms in the second line of (\ref{hhnice}) the situation is subtler.
One may worry that the longitudinal projectors entering them
contain factors $k^2$ in the denominator and apparently
violate the regular form (\ref{regden}). However, we observe that the
combinations in the square brackets in these terms vanish at $k=0$,
$\omega\neq 0$. Besides, they depend on the spatial momentum through
$k^4$. This implies that when the worrisome terms are written as
ratios of polynomials, their numerators are at least proportional to
$k^4$, which cancels all powers of $k$ from the denominator. This
cancellation is in fact guaranteed by the regularity of the propagator
$\langle h_{ij}h_{kl}\rangle$ at $k\to 0$, $\omega$--fixed; this, in
turn, follows from the regular structure of the kinetic term for
$h_{ij}$ 
in this limit, see (\ref{Lnice}).  

The expressions for the propagators are particularly simple for the
choice of the gauge parameters,
\be
\label{supernice}
\s=\frac{1-2\l}{8\m(1-\l)}~,~~~~\xi=-\frac{1-2\l}{2(1-\l)}\;,
\ee
which renders $\PP_1=\PP_2=\PP_s$. Then one obtains,
%\bseq
%\label{propssupernice}
\begin{align}
%\label{NNsupernice}
&\langle N^i(p)N^j(-p)\rangle=4\m\vk^2\bigg[
\frac{2(1-\l)}{1-2\l}\delta_{ij}k^2-k_ik_j\bigg]\PP_s(p)\;,\notag\\
%\label{hhsupernice}
&\langle h_{ij}(p)h_{kl}(-p)\rangle=2\vk^2\bigg[\frac{2\l}{1-2\l}
\delta_{ij}\delta_{kl}
+\delta_{ik}\delta_{jl}+\delta_{il}\delta_{jk}\bigg]
\PP_s(p)\notag\;.
\end{align}
%\eseq
This gauge may be convenient for the actual loop computations in Ho\v rava
gravity. 

%%%%%%%%%%%%%%%%%%%%%%%%%%%%%%%%%
\subsubsection{Fadeev-Popov ghosts}\label{Sec:FP}
%%%%%%%%%%%%%%%%%%%%%%%%%%%%%%%%%

The gauge-fixing procedure must be completed by specifying the
action for the Fad\-de\-ev--Popov ghosts. This is conveniently derived in
the BRST formalism \cite{Becchi:1974md,Tyutin}. 
We follow the presentation of~\cite{Weinberg}. One introduces an
operator ${\bf s}$ transforming the metric and the shift,
\bseq
\label{BRSThN}
\begin{align}
\label{BRSTh}
&{\bf s}h_{ij}=\d_ic_j+\d_jc_i
+\d_ic^kh_{jk}+\d_jc^kh_{ik}+c^k\d_kh_{ij}\;,\\
\label{BRSTN}
&{\bf s}N^i=\dot c^i-N^j\d_jc^i+c^j\d_jN^i\;,
\end{align}
\eseq 
where $c^i(\tau,{\bf x})$ are anticommuting ghost
fields. The transformations (\ref{BRSThN}) are nothing but the
variations of $h_{ij}$ and $N^i$ under infinitesimal diffeomorphisms
with the parameters $c^i$. Supplementing them by the transformation
of the ghosts,
\be
\label{BRSTc}
{\bf s}c^i=c^j\d_jc^i\;,
\ee
it is straightforward to verify that ${\bf s}$ is
nilpotent\footnote{In deriving these identities one uses the graded 
Leibniz rule,
${\bf s} A\cdot B=({\bf s} A)\cdot B+(-1)^{|A|}A\cdot ({\bf s}B)\;,$
where $|A|=0$ ($|A|=1$) for a bosonic (fermionic) field. For example,
$
{\bf s}^2c^i={\bf s} (c^j\d_jc^i)
=({\bf s}c^j)\d_jc^i-c^j\d_j({\bf s}c^i)
$.},
\be
\label{BRSTnil}
{\bf s}^2h_{ij}={\bf s}^2N^{i}={\bf s}^2 c^i=0\;.
\ee
The ghost action is written using the BRST transform of the
gauge-fixing function,
\be
\label{ghostact}
S_{gh}=-\frac{1}{\vk^2}\int \di \tau \di ^2x\;\bar c_i({\bf s}F^i)\;,
\ee 
where we have introduced the antighost $\bar c_i$. Explicitly, upon
integration by parts we obtain,
\be
\label{ghostactd2}
\begin{split}
S_{gh}&=\frac{1}{\vk^2}\int \di \tau \di ^2x\bigg[\dot{\bar c}_i\dot c^i
+\frac{1}{2\s}\D\bar c_i\D c^i
-\frac{1-2\l+2\xi(1-\l)}{2\s}\d_i\bar c_i\D\d_jc^j\\
&-\dot{\bar c}_i\d_jc^iN^j+\dot{\bar c}_ic^j\d_j N^i
-\frac{1}{2\s}\D\d_j\bar c_i
\big(\d_ic^kh_{jk}+\d_jc^kh_{ik}+c^k\d_kh_{ij}\big)\\
&-\frac{\xi}{2\s}\d_i\d_j\d_k\bar c_i
\big(\d_jc^lh_{lk}+\d_kc^lh_{jl}+c^l\d_lh_{jk}\big)
+\frac{\l(1+\xi)}{2\s}\D\d_i\bar c_i\,(2\d_kc^lh_{lk}+c^l\d_l h)\bigg].
\end{split}
\ee
This action is invariant under the anisotropic scaling (\ref{scaling})
with the assignment of zero scaling dimension to the (anti-)ghosts,
\[
%\label{dimghost}
[c^i]=[\bar c_i]=0\;.
\]
Note also that the antighost $\bar c_i$ always 
appear in the action either with
a time derivative or with 3 spatial derivatives acting on it.  
From the first line of (\ref{ghostactd2}) we obtain the propagator,
\be
\label{propcc}
\langle c^i(p)\bar c_j(-p)\rangle=\vk^2\delta_{ij}\PP_1(p)
+\vk^2\hat k_i\hat k_j\big[\PP_2(p)-\PP_1(p)\big]\;.
\ee
It is straightforward to check that this propagator is regular. 
In the
gauge (\ref{supernice}) it diagonalizes,
\[
%\label{supernicecc}
\langle c^i(p)\bar c_j(-p)\rangle=\vk^2\delta_{ij}\PP_s(p)\;.
\]

If we postulate the BRST transformation of the antighost
as\footnote{With this definition the second BRST
  variation of the antighost is non-zero, ${\bf s}^2\bar c_i
=\s\OO_{ij}{\bf s}F^j\neq 0$. It is possible to modify the formalism
in such a way that ${\bf s}^2$ will annihilate all fields, including the
antighost, at the expense of introducing an additional auxiliary
variable. We prefer to avoid this complication which is irrelevant for
our purposes.} 
\be
\label{BRSTbarc}
{\bf s}\bar c_i=\s \OO_{ij} F^j\;,
\ee
the total action composed of the original action, the gauge-fixing
term (\ref{local}) and the action for ghosts is BRST-invariant,
\[
%\label{BRSTsym}
{\bf s}\big[S+S_{gf}+S_{gh}\big]=\frac{1}{\vk^2}\int \di \tau \di ^2x
\big[\s ({\bf s}F^i)\OO_{ij}F^j-({\bf s}\bar c_i)({\bf s}F^i)\big]=0\;,
\]
where we used that the variation of $S$ vanishes 
as the consequence of
gauge
invariance, whereas ${\bf s}^2F^i=0$ due to Eqs.~(\ref{BRSTnil}). 
In other words, the transformations (\ref{BRSThN}),
(\ref{BRSTc}), (\ref{BRSTbarc}) constitute a symmetry of the gauge-fixed
action, reflecting the original gauge invariance. 

This symmetry
explains the following property that at first might appear
surprising. The decomposition of the metric and the shift
(\ref{metrlin}) involves four gauge modes: two transverse vectors
$u_i$, $v_i$ and two longitudinal scalars $B$ and $E$. Adding the
gauge-fixing term to the action makes these modes 
propagating, which naively could give rise to four different pole
structures in the propagators of the metric and the shift. Instead, we
see only two structures associated with the gauge modes,
i.e. $\PP_1$ and $\PP_2$. The reason is that the BRST transformation
connects the gauge modes to the ghost field which contains only one
transverse and one longitudinal component. To see explicitly how this
constrains the propagators, consider the linear part of the 
BRST transformations
(\ref{BRSThN}). They form a symmetry of the quadratic action and hence
leave the two-point Green's functions invariant. 
Let us act with ${\bf s}$ on the
correlators $\langle h_{ij}\,\bar c_k\rangle$ and  
$\langle N^{i}\,\bar c_k\rangle$ which trivially vanish. 
We obtain,
%\bseq
%\label{Ward}
\begin{align}
0={\bf s}\langle h_{ij}\,\bar c_k\rangle
&=\langle (\d_ic_j+\d_j c_i)\,\bar c_k\rangle
+\s\langle h_{ij}\,\OO_{kl}F^l\rangle\notag\\
&=\langle (\d_ic_j+\d_j c_i)\,\bar c_k\rangle
+\frac{1}{2}\langle h_{ij}\,(\d_l h_{kl}-\l\d_kh)\rangle\;,\notag
%\label{Ward1}
\end{align}
where passing to the second line we substituted the explicit form 
(\ref{chisigxi})
of
the gauge-fixing function and used that the correlator
$\langle h_{ij}\,N^k\rangle$ vanishes. Similarly
\[
%\label{Ward2}
0={\bf s}\langle N^i\,\bar c_j\rangle=
\langle \dot c^i\,\bar c_j\rangle+\s\OO_{jk}\langle N^i\,\dot N^k\rangle\;.
\]
%\eseq
These relations imply, in particular, that the poles of the gauge modes must
coincide with the poles of the ghost propagator and thus there can be
at most two gauge-dependent poles.
It is straightforward to verify that the propagators (\ref{propsnice}),
(\ref{propcc}) satisfy the above relations.

\subsection{Theory in 3 spatial dimensions}
\label{sec:4b}
The analysis of the previous section can be easily generalized to
projectable Ho\v rava gravity in spacetime of arbitrary dimension
$(d+1)$. We are going to work out explicitly the case $d=3$, the
lowest dimensionality admitting propagating TT mode. We will not
repeat the details of the derivation, highlighting only the difference
from the $d=2$ case. For the sake of clarity, we keep only
marginal terms in the potential (\ref{Vd3}) omitting the relevant
deformations. The latter do not affect the UV properties of the theory
that are of interest for us.   

The gauge-fixing Lagrangian is still given by the expression
(\ref{local}). However, now the scaling dimension of $F^i$, which
coincides with the dimension of $\dot N^i$, is $5$. Therefore, for the
gauge-fixing term to be scale-invariant, the operator $\OO_{ij}$ must
have dimension\footnote{Recall that for $d=3$ the scaling dimension of
the spacetime measure is $[\di \tau \di ^3x]=-6$.} 
$-4$. Its general form is,
\[
%\label{Od3}
\OO_{ij}=\D^{-1}\big[\delta_{ij}\D+\xi\d_i\d_j\big]^{-1}\;.
\] 
Substituting this into the first line of (\ref{chisigxi}) we obtain
the explicit expression for the gauge-fixing function,
\[
%\label{Fniced3}
F^i=\dot N^i+\frac{1}{2\s} \D^2\d_k h_{ik}
-\frac{\l(1+\xi)}{2\s}\D^2\d_ih+\frac{\xi}{2\s}\D\d_i\d_j\d_kh_{jk}\;.
\]
As in $d=2$, this choice of the gauge-fixing function eliminates the
cross-terms mixing the shift and the metric in the quadratic action.
One combines $\LL_{gf}$ with the quadratic Lagrangian (\ref{S2d3phys}),
sets to zero the coefficients in front of the relevant
deformations,
$\eta=\m_1=\m_2=0$, and flips the sign of $\nu_{4,5}$ in
consequence of the Wick rotation. 
Then, a straightforward calculation
yields the non-zero propagators,
\bseq
\label{propsniced3}
\begin{align}
\label{NNniced3}
\langle N^i(p)N^j(-p) \rangle=&\frac{\vk^2k^2}{\s}
(k^2\delta_{ij}-k_ik_j)\,\PP_1(p)
+\frac{\vk^2(1+\xi)k^2}{\s}k_ik_j\,\PP_2(p)\;,\\
\langle h_{ij}(p)h_{kl}(-p)\rangle=
&2\vk^2
(\delta_{ik}\delta_{jl}+\delta_{il}\delta_{jk})\PP_{tt}(p)
-2\vk^2\delta_{ij}\delta_{kl}\bigg[\PP_{tt}(p)-\frac{1-\l}{1-3\l}
\PP_s(p)\bigg]\notag\\
&-2\vk^2(\delta_{ik}\hat k_j\hat k_l+\delta_{il}\hat k_j\hat k_k
+\delta_{jk}\hat k_i\hat k_l+\delta_{jl}\hat k_i\hat k_k) 
\big[\PP_{tt}(p)-\PP_1(p)\big]\notag\\
&+2\vk^2(\delta_{ij}\hat k_k\hat k_l+\hat k_i\hat k_j\delta_{kl})
\big[\PP_{tt}(p)-\PP_s(p)\big]\notag\\
&+2\vk^2\hat k_i\hat k_j\hat k_k\hat k_l
\bigg[\PP_{tt}(p)+\frac{1-3\l}{1-\l}\PP_s(p)-4\PP_1(p)
+\frac{2\PP_2(p)}{1-\l}\bigg]\;,
\label{hhniced3}
\end{align}
\eseq
where now the pole structures are,
\bseq
\label{PPd3}
\begin{align}
\label{PTT}
&\PP_{tt}=\frac{1}{\omega^2+\nu_5k^6}\;,\\
\label{Psd3}
&\PP_s=\bigg[\omega^2+\frac{(8\nu_4+3\nu_5)(1-\l)}{1-3\l}\,k^6\bigg]^{-1}\;,\\
\label{P1d3}
&\PP_1=\bigg[\omega^2+\frac{k^6}{2\s}\bigg]^{-1}\;,\\
\label{P2d3}
&\PP_2=\bigg[\omega^2+\frac{(1-\l)(1+\xi)}{\s}\,k^6\bigg]^{-1}\;.
\end{align}
\eseq
The first two structures correspond to the physical TT and scalar modes,
cf. Eqs.~(\ref{dispreld3}), whereas the other two are
gauge-dependent. Note that the latter coincide with the
expressions 
(\ref{P12}) up to the substitution $k^4\mapsto k^6$. By inspection
one finds that the propagators (\ref{propsniced3}) satisfy the
regularity conditions (\ref{reg}).

The BRST transformations have the same form as before,
Eqs.~(\ref{BRSThN}), (\ref{BRSTc}), (\ref{BRSTbarc}). This also
applies to the ghost action which is given by (\ref{ghostact}), up to
replacement of the integration measure $\di \tau \di ^2x\mapsto \di \tau
\di ^3x$. We write down explicitly only the quadratic part,
\[
%\label{ghostactd3}
S_{gh}=\frac{1}{\vk^2}\int \di \tau \di ^3x\bigg[\dot{\bar c}_i\dot c^i
-\frac{1}{2\s}\bar c_i\D^3 c^i
+\frac{1-2\l+2\xi(1-\l)}{2\s}\d_i\bar c_i\D^2\d_jc^j+\ldots\bigg]\;,
\]
where dots stand for cubic terms describing interactions of ghosts
with $N^i$ and $h_{ij}$. From this action one reads off the
expression for the ghost propagator which turns out to be the same as
(\ref{propcc}), but with $\PP_{1,2}$ given by Eqs.~(\ref{P1d3}),
(\ref{P2d3}).  

One can use the freedom in the choice of the gauge parameters $\s$ and $\xi$
to simplify the expressions (\ref{propsniced3}). However, unlike the
case $d=2$, it is generically impossible to make all propagators
proportional to each other. This is, of course, due to the presence of
two distinct physical excitations in the theory --- TT and scalar
gravitons --- that in general have different dispersion relations.

%%%%%%%%%%%%%%%%%%%%%%%%%%%%%%%%%
\section{Counterterms}
\label{sec:5}
%%%%%%%%%%%%%%%%%%%%%%%%%%%%%%%%%

In this section we argue that existence of $\s\xi$-gauges where all
propagators are regular implies renormalizability. We  carry out the
derivation for the case $d=2$; the generalization to higher dimensions is
straightforward. 

\subsection{Degree of divergence}
\label{sec:5a}

We work with the total action 
\[
%\label{Stot}
S_{tot}=S+S_{gf}+S_{gh}\;,
\]
where $S$ is given by (\ref{genact}) with the
potential (\ref{Vd2}), the gauge-fixing term $S_{gf}$ corresponds to
the Lagrangian (\ref{local}) and the ghost action $S_{gh}$ has the
form (\ref{ghostact}).
Consider a general diagram appearing in the perturbative
expansion based on this action.
One introduces the notations:
\begin{itemize}
\item $P_{hh}$ --- number of $\langle h_{ij}h_{kl}\rangle$ propagators,
\item $P_{NN}$ --- number of $\langle N^iN^j\rangle$ propagators,
\item $P_{cc}$ --- number of the ghost propagators,
\item $V_{[h]}$ --- number of vertices involving only the $h_{ij}$-fields,
\item $V_{[h]N}$ --- number of vertices with an
  arbitrary number of $h$-legs and a single $N$-leg,
\item $V_{[h]NN}$ --- number of vertices with an
  arbitrary number of $h$-legs and two $N$-legs,
\item $V_{hcc}$ --- number of vertices describing interaction of
  $h_{ij}$ with the ghosts,
\item $V_{Ncc}$ --- number of vertices describing interaction of
  $N^{i}$ with the ghosts,
\item $L$ --- number of loops, i.e. number of independent loop integrals,
\item $l_N$ --- number of external $N$-legs, 
\item $T$ --- number of time-derivatives acting on external legs,
\item $X$ --- number of spatial derivatives acting on external legs.
\end{itemize}
These quantities obey two relations:
\bseq
\label{NLrels}
\begin{align}
\label{Lloops}
&L=P_{hh}+P_{NN}+P_{cc}-V_{[h]}-V_{[h]N}-V_{[h]NN}-V_{hcc}-V_{Ncc}+1\;,\\
\label{Nlines}
&l_N=V_{[h]N}+V_{Ncc}+2V_{[h]NN}-2P_{NN}\;.
\end{align}
\eseq
The first relation follows from the standard reasoning that out of
$(P_{hh}+P_{NN}+P_{cc})$ original integrals over frequencies and momenta
$(V_{[h]}+V_{[h]N}+V_{[h]NN}+V_{hcc}+V_{Ncc}-1)$ 
are removed by the $\delta$-functions
at the vertices (one $\delta$-function remains as an overall factor
multiplying the whole diagram). The second relation is obtained by
counting the $N$-legs. Indeed,
each vertex of the type $V_{[h]N}$ or $V_{Ncc}$
brings one $N$-leg, whereas the vertex $V_{[h]NN}$ brings two; 
every $\langle N^iN^j\rangle$-propagator absorbs two
legs; the remaining $N$-legs are external. 

Next we introduce the superficial degree of divergence $D_{div}$ of the
diagram. This is defined as the scaling power of the diagram under
the simultaneous rescaling of all loop momenta and
frequencies,
\[
%\label{looprescale}
{\bf k}^{(l)}\mapsto b\, {\bf k}^{(l)}~,~~~
\omega^{(l)}\mapsto b^2\, \omega^{(l)}\;,
\]
in the limit $b\to\infty$.
By inspection of the expressions for the propagators and vertices one obtains,
\[
%\label{degreediv}
D_{div}=4L-4P_{hh}-2P_{NN}-4P_{cc}+4V_{[h]}+3V_{[h]N}+2V_{[h]NN}
+4V_{hcc}+3V_{Ncc}-2T-X\;.
\] 
Using (\ref{NLrels}) this reduces to\footnote{
The generalization of this formula for 
$(d+1)$-dimensional Ho\v rava gravity is
$D_{div}=2d-d\cdot T-X-(d-1)l_N$.}
\[
%\label{D2}
D_{div}=4-2T-X-l_N\;.
\]

Let us focus on the diagrams with external $h$-legs only. We see that
$D_{div}$ is negative for diagrams with more than 2 time- or 4
space-derivatives on external legs. Assuming that $D_{div}<0$ implies
convergence of a diagram, one concludes that only
diagrams with at most 2 time- and 4 space-derivatives on the
external lines must be renormalized. These diagrams can be
Taylor expanded\footnote{We assume that the UV divergences have been
  appropriately regulated, e.g. by analytically continuing in the
  dimensionalities of time and space \cite{Anselmi:2007ri}, and that
  possible IR divergences have been removed by introducing an IR
  cutoff.} 
in the external frequencies and momenta, with the
successive terms in the series having lower and lower degree of
divergence. Starting from a certain order the coefficients in the
Taylor expansion become finite, so that
only a few first terms in the series will
require subtraction. The corresponding counterterms are polynomial in
external frequencies and momenta and hence local in position space
space. Again, they have no more than 2 time- or 4 space-derivatives
acting on the metric $h_{ij}$. In other words, their scaling dimension
is less or equal four.
If we further assume that 
the divergent parts of the diagrams
respect the foliation-preserving diffeomorphisms, it follows that 
the counterterms
must have the same form as the terms already present in the action
(\ref{genact}), (\ref{Vd2}). This amounts to renormalizability.   

The above argument contains two important assumptions that we now
scrutinize. First, a generic diagram will contain subdivergences and
thus can diverge despite $D_{div}<0$. Fortunately, as shown in
\cite{Anselmi:2007ri}, the combinatorics of the subtraction procedure
in non-relativistic theories works essentially in the same way as in
the relativistic case, and subdivergences are subtracted by
counterterms introduced at the previous orders of the loop expansion.

Second, even in the absence of subdivergences, the convergence of a
diagram with $D_{div}<0$ is not trivial. Indeed, consider the integral
\[
\int \di \omega^{(1)}\di ^2k^{(1)} \int \prod_{l=2}^L
[\di \omega^{(l)}\di ^2k^{(l)}] \;
f(\{\omega\},\{{\bf k}\})\;,
\]
where we singled out the first loop momentum and suppressed the
dependence on external momenta. Assume for simplicity
that $f$ is a scalar function (in general it can carry tensor 
indices corresponding to the external legs of the diagram).
Because subdivergences are absent,
the inner integral converges and gives a function
$\tilde f\big(\omega^{(1)},k^{(1)}\big)$ which for 
${\bf k}^{(1)}\mapsto b\,{\bf k}^{(1)}$, $\omega^{(1)}\mapsto
b^2\,\omega^{(1)}$ scales as $b^{D_{div}-4}$. 
However, the 
latter 
can
have the form 
\be
\label{spuriousdiv}
\tilde f \sim \big(\omega^{(1)}\big)^{-1+n}
\big(k^{(1)}\big)^{D_{div}-2-2n}~~~
\text{or}~~~
\big(\omega^{(1)}\big)^{-1-n}
\big(k^{(1)}\big)^{D_{div}-2+2n}
~,~~~~n>0\;,
\ee
and the integral over frequency (momentum) will diverge, despite the
fact that the $k$-integral ($\omega$-integral) 
is finite. 
These are precisely the spurious divergences that arise if the
propagators contain irregular contributions discussed in
Sec.~\ref{sec:3}. 
Note that this problem is
absent in Lorentz invariant theories, where the function $\tilde f$ 
can depend only on $\big(\omega^{(1)}\big)^2+\big(k^{(1)}\big)^2$. In
Appendix~\ref{app:B} we prove that spurious divergences
(\ref{spuriousdiv}) do not appear if all propagators have the regular
form (\ref{reg}). In that case $D_{div}<0$ indeed implies convergence
of the diagram.

Finally, we must discuss the gauge invariance of the counterterms. In
the perturbative expansion around flat spacetime, which we have been
considering so far, gauge invariance is actually {\em not} preserved. One way to
proceed would be to exploit the BRST symmetry of the gauge-fixed
action to constrain the structure of counterterms, similar to the
analysis of \cite{Stelle:1976gc}. This approach would require considering
divergent diagrams with external ghost lines (as well as diagrams with
external $N$-legs) and working out their relation to the diagrams
renormalizing the vertices containing only the metric. This should be done order by
order in perturbation theory, and the analysis is very
cumbersome. Instead, we are going to adopt the method of background
effective action\footnote{See
  \cite{DeWitt,Veltman:1975vx,Abbott:1981ke} 
for a pedagogical
  introduction.} where the invariance with respect to the (background)
gauge transformations is manifest.

\subsection{Background-covariant formulation}
\label{sec:5b}

One decomposes the total metric and shift into background
$\bar\gamma_{ij}$, $\bar N^i$ and fluctuations $h_{ij}$, $n^i$:
\[
%\label{decomp}
\gamma_{ij}=\bar\gamma_{ij}+h_{ij}~,~~~~N^i=\bar N^i+n^i\;.
\]
We still have to fix the gauge for the fluctuations. However, one can
do it in the way that explicitly preserves the invariance with respect
to diffeomorphisms acting on the background. This is easily achieved
by covariantizing all formulas of the previous sections. Instead of
(\ref{Onice}), (\ref{chisigxi}) we write,
\[
%\label{Fcovar}
F^i=\bar D_t n^i+\frac{1}{2\s}(\OO^{-1})^{ij}\bar\nabla_k h^k_j
-\frac{\l}{2\s}(\OO^{-1})^{ij}\bar\nabla_j h\;,
\]
where
\[
%\label{timecov}
\bar D_t n^i=\dot n^i-\bar N^k\bar\nabla_k n^i+\bar\nabla_k\bar N^in^k
\]
is the covariant time-derivative and 
\[
%\label{Ocov}
\OO_{ij}=-\big[\bar\D\bar\gamma^{ij}+\xi\bar\nabla^i\bar\nabla^j\big]^{-1}\;.
\]
The covariant derivatives $\bar\nabla_i$ are defined using the
Christoffel connection constructed from $\bar\gamma_{ij}$, all indices
are raised and lowered with $\bar\gamma^{ij}$, $\bar\gamma_{ij}$, and
$h=h_{ij}\bar\gamma^{ij}$. The gauge-fixing action is still given
by the Lagrangian (\ref{local}) that must be integrated over the spacetime with
the covariant measure $\int d\tau d^2x\sqrt{\bar\gamma}$. Similarly, the
ghost action has the form,
\[
%\label{ghostactcovar}
S_{gh}=-\frac{1}{\vk^2}\int \di \tau\di ^2x\sqrt{\bar\gamma}\;\bar c_i({\bf s}F^i)\;,
\]
where BRST transformations of the fields are defined by promoting
all derivatives in (\ref{BRSThN}), (\ref{BRSTc}), (\ref{BRSTbarc})
to be background-covariant. For example, for the
shift we have,
\[
%\label{BRSTncov}
{\bf s}n^i=\bar D_t c^i-n^j\bar\nabla_j c^i+c^j\bar\nabla_j n^i\;.
\]
It is straightforward, even if somewhat tedious, 
to verify that the action of the BRST operator
on $h_{ij}$, $n^i$ and $c^i$ is still nilpotent.

The idea of the background-field method is to take the path integral over the
fluctuations and obtain an effective action for the background
fields. The latter will be automatically invariant with respect to
background gauge transformations\footnote{We assume dimensional regularization, which 
preserves  gauge invariance.}.
 In particular, this holds for its
divergent part that requires renormalization. In other words, in the
background-covariant formulation the counterterms are guaranteed to be
gauge-invariant. 

One should be worried at this point that the gauge-fixing Lagrangian
depends on the background fields in a non-local manner which can
compromise the
locality of the counterterms. To resolve this
issue, we observe that the non-local operator $\OO_{ij}$
actually cancels everywhere in the gauge-fixing action, except the
kinetic term for the shift,
\be
\label{nkin}
S_{n,\,kin}=\int \di \tau\di ^2x\sqrt{\bar\gamma}\frac{\s}{2\vk^2}
\bar D_tn^i\OO_{ij}\bar D_tn^j\;.
\ee 
The latter is cast in the local form by introducing an auxiliary field
$\pi_i$,
\be
\label{nkinpi}
S'_{n,\,kin}=\frac{1}{\vk^2}\int \di \tau \di ^2x\sqrt{\bar\gamma}\bigg[
\frac{1}{2\s}\pi_i(\OO^{-1})^{ij}\pi_j-i\pi_i\bar D_tn^i\bigg]\;.
\ee
Taking the Gaussian path integral over $\pi_i$ reproduces
(\ref{nkin}). Note that we have introduced an imaginary coefficient
in front of the second term in (\ref{nkinpi}) in order to preserve the
positivity of the quadratic term\footnote{Strictly speaking, this
  argument applies in the case $\l<1/2$ when the operator $\OO_{ij}$,
  and hence $(\OO^{-1})^{ij}$, is positive-definite. For $\l>1$ the
  positivity cannot be ensured, which, however, does not affect the
  perturbative considerations, see footnote~\ref{foot:12}.}.
This is not problematic: the imaginary part of (\ref{nkinpi})
is odd when $\pi_i$ changes sign and hence the effective action is real
as it is obtained by integrating over all values of $\pi_i$. Besides,
we notice
that $\pi_i$ enters in the action as a canonically conjugate momentum
for the shift perturbations $n^i$. From this perspective, the presence
of an imaginary part in (\ref{nkinpi}) is not surprising. Indeed, such
imaginary part associated with the canonical form appears even in
ordinary mechanics when the Euclidean action is written in terms of
canonical variables. 

It is instructive to work out how the introduction of $\pi_i$ affects
the measure in the path integral. Let us make a step backward and
recall that the gauge-fixing Lagrangian (\ref{local}) arises as a
result of smearing the gauge-fixing condition $F^i=f^i$ with the
weighting functional
\be
\label{weighting}
\sqrt{\Det \OO_{ij}}\int[\di f^i]\exp\bigg[-\int \di \tau\di ^2x\sqrt{\bar\gamma}
\frac{\s}{2\vk^2}f^i\OO_{ij}f^j\bigg]
\ee
inserted in the partition function of the theory. Notice the square
root of the functional determinant of the operator $\OO_{ij}$ 
in the prefactor which ensures the
correct normalization. Thus, before 
introducing $\pi_i$ the partition function has the form,
\[
%\label{partit1}
Z=\sqrt{\Det \OO_{ij}}\int[\di n^j\di h_{kl}\di c^m\di \bar c_n]
\exp\big[-(S_{n,\,kin}+\ldots)\big]\;,
\]
where ellipsis stands for the local contributions in the action. The
introduction of $\pi_i$ not only makes the action local, but also
absorbs the determinant from the prefactor. This follows
from the relations,
\begin{align}
&\e^{-S_{n,\,kin}[n^i]}=\sqrt{\Det(\OO^{-1})^{ij}}
\int [\di\pi_j]\e^{-S'_{n,\,kin}[\pi_j,n^i]}\;,\notag\\
&\sqrt{\Det\OO_{ij}}\sqrt{\Det(\OO^{-1})^{ij}}=1\;.\notag
\end{align}
Thus, we arrive to the final expression for the partition function,
\[
%\label{partit2}
Z=\int[\di \pi_i\di n^j\di h_{kl}\di c^m\di \bar c_n]
\exp\big[-(S'_{n,\,kin}+\ldots)\big]\;.
\]
Curiously, the introduction of $\pi_i$ makes the integration measure
in the path integral flat, which further supports the identification
of $\pi_i$ as the canonically
conjugate momentum to~$n^i$.

Last but not least, we have to check that the introduction of $\pi_i$ does
not spoil the regular structure of the propagators. It is sufficient
to perform this analysis for the flat background,
$\bar\gamma_{ij}=\delta_{ij}$, $\bar N^i=0$, as locally any background
can be brought to this form by a coordinate choice and our question deals with 
the local properties of the propagators. From the quadratic
Lagrangian in the $\pi n$-sector,
\[
%\label{pinquadr}
\LL_{2,\,\pi n}^{d=2}=\frac{1}{2\vk^2}\bigg[-\frac{1}{\s}
\pi_i(\D\delta_{ij}+\xi\d_i\d_j)\pi_j-2i\pi_k\dot n^k
+\frac{(\d_in^j)^2}{2}+\bigg(\frac{1}{2}-\l\bigg)(\d_in^i)^2\bigg]\;,
\]
we find that the $\langle n^in^j\rangle$ propagator is not modified
and is given by (\ref{NNnice}), whereas
%\bseq
%\label{piprops}
\begin{align}
%\label{pinprop}
&\langle \pi_i(p)n^j(-p)\rangle=\vk^2\omega\delta_{ij}\PP_1(p)
+\vk^2\omega\hat k_i\hat k_j\big[\PP_{2}(p)-\PP_1(p)\big]\;,\notag\\
%\label{pipiprop}
&\langle \pi_i(p)\pi_j(-p)\rangle=\frac{\vk^2}{2}(k^2\delta_{ij}-k_ik_j)\PP_1(p)
+\vk^2(1-\l)k_i k_j\PP_{2}(p)\;.\notag
\end{align}
%\eseq
These are compatible with the regular form (\ref{reg}) for the scaling
dimension\footnote{The fact that the scaling dimension of $\pi_i$ is
  equal to 1 follows from the way it enters into the action multiplied
  by the time-derivative of the shift. This applies to
  Ho\v rava gravity in any number of spacetime dimensions.} $[\pi_i]=1$.
As a consequence, the reasoning of Sec.~\ref{sec:5a} goes through 
essentially unchanged with the field $\pi_i$ included into consideration.

To sum up, we have formulated a manifestly background-covariant
gauge-fixing procedure. Combined with the results of Sec.~\ref{sec:5a}
about the scaling dimension of possible divergences, it
implies that
one loop counterterms in the effective action have the same
structure as the terms in the bare Lagrangian. A careful treatment
\cite{Barvinsky:2017zlx} 
shows that this also holds at higher loops (see
\cite{Abbott:1980hw,Barvinsky:1988ds,Grassi:1995wr,Anselmi:2013kba}
for earlier arguments). Therefore, the theory is renormalizable.

%%%%%%%%%%%%%%%%%%%%%%%%%%%%%%%%%
\section{Non-projectable model}
\label{sec:6}
%%%%%%%%%%%%%%%%%%%%%%%%%%%%%%%%%

We now consider the non-projectable Ho\v rava gravity. Again, for
illustration of the general situation we take the model in
$(2+1)$-dimensions \cite{Sotiriou:2011dr} which is technically 
much simpler than
its $(3+1)$-dimensional counterpart. Upon using integration by parts
and the fact that in two dimensions the Riemann tensor is expressed in
terms of the scalar curvature, one finds that the potential contains
10 inequivalent terms,
\[
\begin{split}
%\label{Vd2NP}
{\cal V}=&2\Lambda-\eta R-\a a_ia^i+\m R^2+\rho_1\D R+\rho_2 R a_ia^i
\\
&+\rho_3 (a_ia^i)^2+\rho_4a_ia^i\nabla_ja^j+\rho_5(\nabla_j a^j)^2
+\rho_6\nabla_ia_j\nabla^ia^j\;,
\end{split}
\]
where
\[
%\label{ai}
a_i=\frac{\d_iN}{N}
\]
is the combination of the lapse and its derivative which is invariant
under the reparameterizations of time, see
Eqs.~(\ref{Ftrans}). Expanding around flat spacetime\footnote{For the
  flat spacetime to be a solution, we assume that the cosmological
  constant $\Lambda$ is tuned to zero.}, one obtains the
quadratic action,
\[
%\label{S2NP}
\begin{split}
&S_2^{d=2,\, np.}=\frac{1}{2\vk^2}\int \di t\di ^2x\bigg[
\!-\frac{1}{2}(\dot v_i-u_i)\D(\dot v_i-u_i)
+\frac{\dot\psi^2}{4}+\frac{1}{4}(\dot E-2\D B)^2
\\
&~~~~~-\frac{\l}{4}(\dot\psi+\dot E-2\D B)^2-\eta\phi\D\psi-\a\phi\D\phi
-\m(\D\psi)^2+\rho_1\phi\D^2\psi-(\rho_5+\rho_6)\phi\D^2\phi\bigg],
\end{split}
\]
where we used the helicity decomposition (\ref{metrlin}) and
introduced the fluctuation of the lapse $\phi\equiv N-1$. This action
propagates a single scalar degree of freedom with the dispersion
relation,
\[
%\label{disprelNP}
\omega^2=\left(\frac{1-\l}{1-2\l}\right)
\frac{\eta^2k^2+(4\a\m+2\eta\rho_1)k^4
+\big(\rho_1^2-4\m(\rho_5+\rho_6)\big)k^6}{\a-(\rho_5+\rho_6)k^2}\;.
\]
In contrast to the projectable case, this dispersion relation is
linear at low $k$,
\[
%\label{disprelNPlowk}
\omega^2=\left(\frac{1-\l}{1-2\l}\right)\frac{\eta^2}{\a}\; k^2\;.
\]
At large momenta it respects the anisotropic scaling (\ref{scaling}),
\[
%\label{disprelNPhighk}
\omega^2=\frac{1-\l}{1-2\l}\bigg(4\m-\frac{\rho_1^2}{\rho_5+\rho_6}\bigg)
\; k^4\;.
\]
The mode has positive energy and is stable at all momenta for 
an appropriate choice of parameters. In particular, the following
necessary conditions must be satisfied:
$\vk^2>0$, 
$\l<1/2$ or $\l>1$, $\a>0$, $(\rho_5+\rho_6)<0$ and
$4\m>\rho_1^2/(\rho_5+\rho_6)$. 

Let us focus on the UV behavior of the model. Accordingly, we retain
only marginal terms in the action which amounts to setting
$\eta=\a=0$. Next, one performs the Wick rotation and introduces the
gauge-fixing term (\ref{local}) with the gauge-fixing function
(\ref{chisigxi}) and $\OO_{ij}$ defined in (\ref{Onice}). After a
somewhat lengthy, but straightforward calculation one finds that the
propagators of the shift and the metric have the form
(\ref{propsnice}), where $\PP_1$, $\PP_2$ are still given by
(\ref{P12}), while
\[
%\label{PsNP}
\PP_s(p)=\bigg[\omega^2+\frac{1-\l}{1-2\l}
\bigg(4\m-\frac{\rho_1^2}{\rho_5+\rho_6}\bigg)\,k^4\bigg]^{-1}\;.
\]
Other non-vanishing propagators are 
\bseq
\label{phiprops}
\begin{align}
\label{phiphiprop}
&\langle\phi(p)\phi(-p)\rangle=
\frac{\vk^2(1-\l)\rho_1^2}{(1-2\l)(\rho_5+\rho_6)^2}\PP_s(p)
+\frac{\vk^2}{(\rho_5+\rho_6)k^4}\;,\\
\label{phihprop}
&\langle\phi(p)h_{ij}(-p)\rangle=
\frac{2\vk^2(1-\l)\rho_1}{(1-2\l)(\rho_5+\rho_6)}\delta_{ij}\PP_s(p)
-\frac{2\vk^2\rho_1k_ik_j}{(\rho_5+\rho_6)k^2}\,\PP_s(p)\;.
\end{align}
\eseq
Clearly, the last terms in (\ref{phiphiprop}), (\ref{phihprop})
violate the regularity condition. Though these contributions have been derived
within a particular family of gauges, we believe they cannot be
removed by any gauge choice. They correspond to the
instantaneous interaction present in the theory
\cite{Blas:2010hb,Blas:2011ni}. We conclude that the correlators of
the lapse contain genuinely non-local terms whose 
contributions to the loop diagrams must be carefully analyzed to  
establish or disprove the renormalizability of
the theory.

%%%%%%%%%%%%%%%%%%%%%%%%%%%%%%%%%
\section{Conclusions}
\label{sec:7}
%%%%%%%%%%%%%%%%%%%%%%%%%%%%%%%%%

In this paper we have demonstrated renormalizability of the
\emph{projectable} version of Ho\v rava gravity. Though for concreteness we focused on the
models in $d=2$ and $3$ spatial dimensions, our analysis is completely
general and applies to Ho\v rava gravity in any dimensionality. Thus, for $d\geq
3$ projectable Ho\v rava gravity presents the first example of unitary
renormalizable quantum field theory of gravitation with dynamical
transverse-traceless excitations (gravitons). The key element of our
argument is the choice of gauge-fixing which ensures the right scaling
properties of the propagators and their uniform falloff at large
frequencies and momenta. The latter property is essential to guarantee
locality of the singularities of the propagators in position
space. An unusual feature of our approach is 
that it
involves a gauge-fixing term which is non-local in space. 
We showed how this
non-locality can be resolved by introduction of an auxiliary field
resulting in a local gauge-fixed action.  

We restricted the analysis to pure gravity theories. We now argue that
the renormalizability is preserved upon inclusion of
matter obeying the Lifshitz scaling (\ref{scaling}) in the UV. Indeed,
the derivation of Sec.~\ref{sec:5} relies only on the scaling
properties of the propagators and vertices, regular form of the
propagators, and invariance under foliation-preserving
diffeomorphisms. These properties are clearly satisfied by Lifshitz
scalars and fermions, and in general, by any Lifshitz matter with only
global internal symmetries. 
For gauge theories one should use an
appropriate gauge-fixing that leads to regular propagators. It turns
out that this gauge-fixing is non-local, in complete analogy with what
we have found for gravity. For example, for a gauge field $({\cal
  A}_0,{\cal A}_i)$ in $d=2$ with the quadratic Euclidean Lagrangian,
\be
\label{LagrA}
{\cal L}_A=\frac{1}{2}{\cal F}_{0i}{\cal F}^{0i}
-\frac{\m_A}{4}{\cal F}_{ij}\D {\cal F}^{ij}~,~~~~~~
{\cal F}_{0i}=\dot{\cal A}_i-\d_i{\cal A}_0~,~~
{\cal F}_{ij}=\d_i{\cal A}_j-\d_j{\cal A}_i\;,
\ee
the suitable gauge-fixing term is
\[
{\cal L}_{A,\,gf}=-\frac{\s_A}{2}
\bigg(\dot{\cal A}_0+\frac{1}{\s_A}\D\d_i{\cal A}_i\bigg)\frac{1}{\D}
\bigg(\dot{\cal A}_0+\frac{1}{\s_A}\D\d_j{\cal A}_j\bigg)\;.
\]
It is chosen in the way to cancel the cross-terms mixing ${\cal A}_0$
and ${\cal A}_i$. For $\s_A=1/\m_A$ the quadratic action diagonalizes
completely and the propagators are particularly simple,
\[
\langle {\cal A}_0 {\cal
  A}_0\rangle=\frac{\m_Ak^2}{\omega^2+\m_Ak^4}~,~~~~
\langle {\cal A}_i {\cal
  A}_j\rangle=\frac{\delta_{ij}}{\omega^2+\m_Ak^4}\;.
\]
These are regular and correspond to the dimensions $[{\cal A}_0]=1$,
$[{\cal A}_i]=0$ dictated by the anisotropic scale-invariance of the
Lagrangian (\ref{LagrA}). Furthermore, the non-locality of
the gauge-fixing term gets resolved by introducing a canonically
conjugate momentum for ${\cal A}_0$, as done in Sec.~\ref{sec:5b}
for gravity. We conclude that the projectable Ho\v rava gravity can be
coupled also to gauge fields with Lifshitz scaling without spoiling
renormalizability. 

Given renormalizability, the next obvious question is the
renormalization group (RG) behavior of the coupling constants. Only if
the RG flow has a weakly coupled UV fixed point, can the theory be
considered as UV-complete. If instead it turns out that the running
couplings develop a Landau pole, the theory will still require
embedding into a broader framework. Our results imply that
the running of the couplings is logarithmic, and thus the need for
such embedding, if any, will be postponed to exponentially high
energy scales. 

Another issue that acquires new importance in view of our findings is
the fate of the instability at low momenta present in
$(3+1)$-dimensional 
projectable Ho\v rava gravity. It would be interesting to understand if this
instability can be cut off by non-linear terms in the Lagrangian and,
if yes, determine the structure of the ground state. The fact that the
instability is developed by the modes with non-zero spatial momenta
suggests that the putative ground state will break translational
invariance.

For the non-projectable Ho\v rava gravity our gauge-fixing procedure is
not sufficient to establish renormalizability. We have found that it still
leaves certain contributions in the propagators of the lapse that do
not fall off with frequency. These contributions cannot be removed by
any gauge choice and are related to the physical instantaneous
interaction present in the theory. In position space they lead to
singularities in the propagators that are non-local in space. It will
be crucial to work out the implications of this non-locality for the
perturbative expansion in order to establish (or disprove)
renormalizability of the theory. We leave this study for future.

\paragraph{Acknowledgments} We are indebted to Enrique \'Alvarez, 
Dmitry Levkov, Valery
Rubakov and Arkady Vainshtein for stimulating discussions. We thank
Nikolai Krasnikov and Mikhail Shaposhnikov for encouraging
interest. 
A.B. is grateful for hospitality and support of Theory
Division, CERN,
and Pacific Institute of Theoretical Physics, UBC. The work of A.B. was also
supported by the RFBR grant No. 14-02-01173 and by the Tomsk State
University Competitiveness Improvement Program. 
M. H-V. wishes to thank the Institut de Th\'eorie des Ph\'enom\`enes
Physiques at EPFL and the Physikalisches Institut at the University of
Freiburg for kind hospitality and support. 
The work of M. H-V. has been supported by the European Union FP7 ITN
INVISIBLES (Marie Curie Actions, PITN-GA-2011-289442), COST action
MP1210 (The String Theory Universe) and by the Spanish MINECO Centro
de Excelencia Severo Ochoa Programme under grant SEV-2012-0249. 
S.S. thanks Kavli Institute for Theoretical Physics at Santa
Barbara for hospitality during the initial stage of this project (the
visit to KITP was supported in part by the National Science Foundation
under Grant NSF PHY11-25915). 
The work of S.S. is supported by the
Swiss National Science Foundation.
C.S. gratefully acknowledges the kind hospitality and support of the CERN Theory
Division and the Instituto de F\'isica Te\'orica UAM-CSIC while part of this
work was done.

\appendix
%%%%%%%%%%%%%%%%%%%%%%%%%%%%%%%%%
\section{A tensor identity}
\label{app:A}
%%%%%%%%%%%%%%%%%%%%%%%%%%%%%%%%%

In dimension $d=2$ any tensor antisymmetric in 3 or more
indices is identically equal to zero. Consider the combination 
$\hat k_m\delta^i_j\delta^k_l$, where $\hat k_m$ is a unit vector. By
antisymmetrizing it over the lower indices and contracting with $\hat
k^m$ we obtain,
\[
%\label{firstident}
0=\hat k^m \hat k_{[m}\delta^i_j\delta^{k}_{l]}=
\delta^i_j\delta^{k}_{l}-\delta^i_l\delta^{k}_{j}
+\hat k^k\hat k_j\delta^i_l-\hat k^i\hat k_j\delta^k_l
+\hat k^i\hat k_l\delta^k_j-\hat k^k\hat k_l\delta^i_j\;.
\]
Next, we lower the indices and symmetrize in $i$ and $j$. This yields, 
\be
\label{identd2}
2\delta_{ij}\delta_{kl}-\delta_{ik}\delta_{jl}-\delta_{il}\delta_{jk}
-2\delta_{ij}\hat k_k\hat k_l-2\hat k_i\hat k_j\delta_{kl}
+\hat k_i\hat k_k\delta_{jl}+\hat k_j\hat k_k\delta_{il}
+\hat k_i\hat k_l\delta_{jk}+\hat k_j\hat k_l\delta_{ik}=0\;.
\ee
%This identity is used in the main text.

%%%%%%%%%%%%%%%%%%%%%%%%%%%%%%%%%
\section{Convergence of loop integrals}
\label{app:B}
%%%%%%%%%%%%%%%%%%%%%%%%%%%%%%%%%

In this Appendix we use the conventions and notations of
Sec.~\ref{sec:5a}. We prove 
the following statement:

{\it Consider a diagram with $L$ loops and $D_{div}<0$. Assume that
  all propagators in the diagram are regular in the sense (\ref{reg})
  and that if the momentum and frequency in any of the propagators are
  frozen, the integral over remaining
  momenta and frequencies 
converges (i.e. subdivergences are absent).
Then the whole diagram converges.\footnote{Here we are talking about
  convergence in the UV. Infrared divergences present a separate
  issue and must be regulated by an IR cutoff, see below.}
 }

\underline{Proof:} Suppressing the external momenta, the
expression for the diagram takes the form,
\be
\label{ID}
I_D=\int \prod_{l=1}^L [\di\omega^{(l)}\di^2k^{(l)}]\;
{\cal F}_n(\{\omega\},\{{\bf k}\})\;
\prod_{m=1}^M \Big[\big(A_m\Omega^{(m)}\big)^2+B_m\big(K^{(m)}\big)^4\Big]^{-1}\;,
\ee
where ${\cal F}_n$ is a polynomial of scaling 
degree $n$; $\Omega^{(m)}$,
$K^{(m)}$ are linear combinations of loop frequencies and momenta;
and the coefficients $A_m$, $B_m$ are strictly positive, $A_m,B_m>0$. 
The parameters in
(\ref{ID}) satisfy,
\be
\label{LMD}
4L+n-4M=D_{div}\;.
\ee
It is convenient to
transform (\ref{ID}) into the Schwinger-type representation using 
\[
x^{-1}=\int_0^{\infty}\di s\;\e^{-sx}\;.
\]
This yields,
\be
\label{ID1}
I_D=\int_0^\infty \prod_{m=1}^M \di s_m \;G(\{s\})\;,
\ee
where
\be
\label{Gs}
G(\{s\})=\int \prod_{l=1}^L [\di\omega^{(l)}\di^2k^{(l)}]\;
{\cal F}_n(\{\omega\},\{{\bf k}\})\;
\prod_{m=1}^M 
\exp
\bigg\{-s_m\Big[A_m\big(\Omega^{(m)}\big)^2+B_m\big(K^{(m)}\big)^4\Big]\bigg\}\;.
\ee
Let us introduce the parameterization,
\[
%\label{xparam}
s_m=\bar s\, x_m~,~~~~\sum_{m=1}^M x_m=1\;,
\] 
Using the scaling properties of the integrand in (\ref{Gs}) we
obtain,
\[
%\label{Gs1}
G(\{s\})=({\bar s})^{-L-n/4} G(\{x\})\;.
\]
Substituting into (\ref{ID1}) and introducing UV and IR regulators
$\bar s_0$, $\bar s_1$ we obtain,
\[
%\label{IDreg}
I_D^{reg}=\int_{\bar s_0}^{\bar s_1} \di\bar s\; (\bar s)^{-D_{div}/4-1}~ \tilde I\;,
\]
where 
\[
\tilde I=\int
\prod_{m=1}^M \di x_m\;\delta\Big(\sum_{m=1}^M x_m-1\Big) G(\{x\})\;,
\]
is the integral over ``angles'' and 
we have used the relation (\ref{LMD}) to write the overall power of
$\bar s$. 
If $\tilde I$ 
converges, the UV regulator $\bar s_0$ can
be removed, i.e. the diagram is UV finite 
(recall that $D_{div}$ is negative\footnote{Clearly, in this
case the diagram exhibits an IR divergence at the upper end
of integration over~$\bar s$.}).

Thus we have to analyze the convergence of $\tilde I$. By inspection of
(\ref{Gs}) we see that $G(\{x\})$ can have singularities only at the
points where some $x_m$ vanishes --- otherwise the integral over
frequencies and momenta in (\ref{Gs}) is damped by the
exponentials\footnote{Note that this step in the
argument relies on the strict positivity of $A_m$ and $B_m$.}. 
The most dangerous singularity occurs when all $x$'s, except one,
tend to
zero, 
\[
%\label{xdang}
x_1\approx 1~,~~~~x_m\to 0\;,~m=2,\ldots, M\;.
\] 
The integral over ``angles'' in the $\epsilon$-vicinity of this point takes the
form,
\be
\label{Ieps}
\begin{split}
&\tilde I_\epsilon\approx\int_0^\epsilon 
\prod_{m=2}^M \di x_m\; G(\{x\})=
\int \di\omega^{(1)}\di^2k^{(1)}
\exp
\Big\{-\Big[A_1\big(\omega^{(1)}\big)^2+B_1\big(k^{(1)}\big)^4\Big]\Big\}\\
&\times
\int 
\prod_{l=2}^L [\di\omega^{(l)}\di^2k^{(l)}]\;
{\cal F}_n(\{\omega\},\{{\bf k}\})\;
\prod_{m=2}^M 
\frac{
1-\exp
\Big\{-\epsilon
\Big[A_m\big(\Omega^{(m)}\big)^2+B_m\big(K^{(m)}\big)^4\Big]\Big\}}
{A_m\big(\Omega^{(m)}\big)^2+B_m\big(K^{(m)}\big)^4}\;,
\end{split}
\ee 
where, without loss of generality, we have identified the frequency and momentum
flowing through the first propagator with $\omega^{(1)}$, $k^{(1)}$.  
The integral in the second line of (\ref{Ieps}) converges. Indeed,
it is free from IR divergences, because the integrand is regular
at $\omega^{(l)}, k^{(l)}\to 0$, whereas a UV divergence is absent by 
assumption. Furthermore, the integrand can be bounded
by rational functions, so the total integral grows at most polynomially 
in $\omega^{(1)}$ and
$k^{(1)}$. Then the integral over $\omega^{(1)}$, $k^{(1)}$
converges as well\footnote{Again, it is important that both $A_1$ and $B_1$ are
strictly positive.}.
We conclude that 
$\tilde I_\epsilon$, and hence $\tilde I$, converges, which completes the proof.


\begin{thebibliography}{99}

\bibitem{Polchinski}
J.~Polchinski, ``String Theory,'' Cambridge University Press 1998.

%\cite{Stelle:1976gc}
\bibitem{Stelle:1976gc} 
  K.~S.~Stelle,
  %``Renormalization of Higher Derivative Quantum Gravity,''
  Phys.\ Rev.\ D {\bf 16}, 953 (1977).
  %%CITATION = PHRVA,D16,953;%%
  %1141 citations counted in INSPIRE as of 14 Nov 2015

%\cite{Fradkin:1981hx}
\bibitem{Fradkin:1981hx} 
  E.~S.~Fradkin and A.~A.~Tseytlin,
  %``Renormalizable Asymptotically Free Quantum Theory of Gravity,''
  Phys.\ Lett.\ B {\bf 104}, 377 (1981); 
  %%CITATION = PHLTA,B104,377;%%
  %135 citations counted in INSPIRE as of 14 Nov 2015
%\cite{Fradkin:1981iu}
%\bibitem{Fradkin:1981iu} 
%  E.~S.~Fradkin and A.~A.~Tseytlin,
  %``Renormalizable asymptotically free quantum theory of gravity,''
  Nucl.\ Phys.\ B {\bf 201}, 469 (1982).
  %%CITATION = NUPHA,B201,469;%%
  %343 citations counted in INSPIRE as of 14 Nov 2015

%\cite{Avramidi:1985ki}
\bibitem{Avramidi:1985ki} 
  I.~G.~Avramidi and A.~O.~Barvinsky,
  %``Asymptotic Freedom In Higher Derivative Quantum Gravity,''
  Phys.\ Lett.\ B {\bf 159}, 269 (1985).
  %%CITATION = PHLTA,B159,269;%%
  %147 citations counted in INSPIRE as of 14 Nov 2015

%\cite{Salvio:2014soa}
\bibitem{Salvio:2014soa} 
  A.~Salvio and A.~Strumia,
  %``Agravity,''
  JHEP {\bf 1406}, 080 (2014)
  [arXiv:1403.4226 [hep-ph]].
  %%CITATION = ARXIV:1403.4226;%%
  %47 citations counted in INSPIRE as of 14 Nov 2015

%\cite{Einhorn:2014gfa}
\bibitem{Einhorn:2014gfa} 
  M.~B.~Einhorn and D.~R.~T.~Jones,
  %``Naturalness and Dimensional Transmutation in Classically Scale-Invariant Gravity,''
  JHEP {\bf 1503}, 047 (2015)
  [arXiv:1410.8513 [hep-th]].
  %%CITATION = ARXIV:1410.8513;%%
  %12 citations counted in INSPIRE as of 14 Nov 2015

%\cite{Horava:2008ih}
\bibitem{Horava:2008ih} 
  P.~Horava,
  %``Membranes at Quantum Criticality,''
  JHEP {\bf 0903}, 020 (2009)
  [arXiv:0812.4287 [hep-th]].
  %%CITATION = ARXIV:0812.4287;%%
  %477 citations counted in INSPIRE as of 14 Nov 2015

%\cite{Horava:2009uw}
\bibitem{Horava:2009uw} 
  P.~Horava,
  %``Quantum Gravity at a Lifshitz Point,''
  Phys.\ Rev.\ D {\bf 79}, 084008 (2009)
  [arXiv:0901.3775 [hep-th]].
  %%CITATION = ARXIV:0901.3775;%%
  %1189 citations counted in INSPIRE as of 14 Nov 2015

%\cite{Mukohyama:2010xz}
\bibitem{Mukohyama:2010xz} 
  S.~Mukohyama,
  %``Horava-Lifshitz Cosmology: A Review,''
  Class.\ Quant.\ Grav.\  {\bf 27}, 223101 (2010)
  [arXiv:1007.5199 [hep-th]].
  %%CITATION = ARXIV:1007.5199;%%
  %144 citations counted in INSPIRE as of 14 Nov 2015

%\cite{Sotiriou:2010wn}
\bibitem{Sotiriou:2010wn} 
  T.~P.~Sotiriou,
  %``Horava-Lifshitz gravity: a status report,''
  J.\ Phys.\ Conf.\ Ser.\  {\bf 283}, 012034 (2011)
  [arXiv:1010.3218 [hep-th]].
  %%CITATION = ARXIV:1010.3218;%%
  %110 citations counted in INSPIRE as of 14 Nov 2015

%\cite{Blas:2009qj}
\bibitem{Blas:2009qj} 
  D.~Blas, O.~Pujolas and S.~Sibiryakov,
  %``Consistent Extension of Horava Gravity,''
  Phys.\ Rev.\ Lett.\  {\bf 104}, 181302 (2010)
  [arXiv:0909.3525 [hep-th]].
  %%CITATION = ARXIV:0909.3525;%%
  %290 citations counted in INSPIRE as of 14 Nov 2015

%\cite{Blas:2010hb}
\bibitem{Blas:2010hb} 
  D.~Blas, O.~Pujolas and S.~Sibiryakov,
  %``Models of non-relativistic quantum gravity: The Good, the bad and the healthy,''
  JHEP {\bf 1104}, 018 (2011)
  [arXiv:1007.3503 [hep-th]].
  %%CITATION = ARXIV:1007.3503;%%
  %138 citations counted in INSPIRE as of 14 Nov 2015

%\cite{Blas:2011en}
\bibitem{Blas:2011en} 
  D.~Blas and S.~Sibiryakov,
  %``Technically natural dark energy from Lorentz breaking,''
  JCAP {\bf 1107}, 026 (2011)
  [arXiv:1104.3579 [hep-th]];
  %%CITATION = ARXIV:1104.3579;%%
  %20 citations counted in INSPIRE as of 14 Nov 2015
%\cite{Audren:2013dwa}
%\bibitem{Audren:2013dwa} 
  B.~Audren, D.~Blas, J.~Lesgourgues and S.~Sibiryakov,
  %``Cosmological constraints on Lorentz violating dark energy,''
  JCAP {\bf 1308}, 039 (2013)
  [arXiv:1305.0009 [astro-ph.CO]].
  %%CITATION = ARXIV:1305.0009;%%
  %19 citations counted in INSPIRE as of 14 Nov 2015

%\cite{Blas:2012vn}
\bibitem{Blas:2012vn} 
  D.~Blas, M.~M.~Ivanov and S.~Sibiryakov,
  %``Testing Lorentz invariance of dark matter,''
  JCAP {\bf 1210}, 057 (2012)
  [arXiv:1209.0464 [astro-ph.CO]];
  %%CITATION = ARXIV:1209.0464;%%
  %20 citations counted in INSPIRE as of 14 Nov 2015
%\cite{Audren:2014hza}
%\bibitem{Audren:2014hza}
  B.~Audren, D.~Blas, M.~M.~Ivanov, J.~Lesgourgues and S.~Sibiryakov,
  %``Cosmological constraints on deviations from Lorentz invariance in gravity and dark matter,''
  JCAP {\bf 1503} (2015) 03,  016
  [arXiv:1410.6514 [astro-ph.CO]].
  %%CITATION = ARXIV:1410.6514;%%
  %7 citations counted in INSPIRE as of 14 Nov 2015

%\cite{Yagi:2013qpa}
\bibitem{Yagi:2013qpa} 
  K.~Yagi, D.~Blas, N.~Yunes and E.~Barausse,
  %``Strong Binary Pulsar Constraints on Lorentz Violation in Gravity,''
  Phys.\ Rev.\ Lett.\  {\bf 112}, no. 16, 161101 (2014)
  [arXiv:1307.6219 [gr-qc]];
  %%CITATION = ARXIV:1307.6219;%%
  %33 citations counted in INSPIRE as of 14 Nov 2015
%\cite{Yagi:2013ava}
%\bibitem{Yagi:2013ava} 
  K.~Yagi, D.~Blas, E.~Barausse and N.~Yunes,
  %``Constraints on Einstein-Aether theory and Ho\v rava gravity from binary pulsar observations,''
  Phys.\ Rev.\ D {\bf 89}, no. 8, 084067 (2014)
  [Phys.\ Rev.\ D {\bf 90}, no. 6, 069902 (2014)]
  [Phys.\ Rev.\ D {\bf 90}, no. 6, 069901 (2014)]
  [arXiv:1311.7144 [gr-qc]].
  %%CITATION = ARXIV:1311.7144;%%
  %41 citations counted in INSPIRE as of 14 Nov 2015

%\cite{Blas:2014aca}
\bibitem{Blas:2014aca} 
  D.~Blas and E.~Lim,
  %``Phenomenology of theories of gravity without Lorentz invariance: the preferred frame case,''
  Int.\ J.\ Mod.\ Phys.\ D {\bf 23}, no. 13, 1443009 (2015)
  [arXiv:1412.4828 [gr-qc]].
  %%CITATION = ARXIV:1412.4828;%%
  %7 citations counted in INSPIRE as of 14 Nov 2015

%\cite{GrootNibbelink:2004za}
\bibitem{GrootNibbelink:2004za} 
  S.~Groot Nibbelink and M.~Pospelov,
  %``Lorentz violation in supersymmetric field theories,''
  Phys.\ Rev.\ Lett.\  {\bf 94}, 081601 (2005)
  [hep-ph/0404271];
  %%CITATION = HEP-PH/0404271;%%
  %116 citations counted in INSPIRE as of 14 Nov 2015
%\cite{Bolokhov:2005cj}
%\bibitem{Bolokhov:2005cj} 
  P.~A.~Bolokhov, S.~Groot Nibbelink and M.~Pospelov,
  %``Lorentz violating supersymmetric quantum electrodynamics,''
  Phys.\ Rev.\ D {\bf 72}, 015013 (2005)
  [hep-ph/0505029].
  %%CITATION = HEP-PH/0505029;%%
  %79 citations counted in INSPIRE as of 14 Nov 2015

%\cite{Pujolas:2011sk}
\bibitem{Pujolas:2011sk} 
  O.~Pujolas and S.~Sibiryakov,
  %``Supersymmetric Aether,''
  JHEP {\bf 1201}, 062 (2012)
  [arXiv:1109.4495 [hep-th]].
  %%CITATION = ARXIV:1109.4495;%%
  %19 citations counted in INSPIRE as of 14 Nov 2015

%\cite{Pospelov:2010mp}
\bibitem{Pospelov:2010mp} 
  M.~Pospelov and Y.~Shang,
  %``On Lorentz violation in Horava-Lifshitz type theories,''
  Phys.\ Rev.\ D {\bf 85}, 105001 (2012)
  [arXiv:1010.5249 [hep-th]].
  %%CITATION = ARXIV:1010.5249;%%
  %51 citations counted in INSPIRE as of 14 Nov 2015

%\cite{Bednik:2013nxa}
\bibitem{Bednik:2013nxa} 
  G.~Bednik, O.~Pujol\`as and S.~Sibiryakov,
  %``Emergent Lorentz invariance from Strong Dynamics: Holographic examples,''
  JHEP {\bf 1311}, 064 (2013)
  [arXiv:1305.0011 [hep-th]];
  %%CITATION = ARXIV:1305.0011;%%
  %14 citations counted in INSPIRE as of 14 Nov 2015
%\cite{Kharuk:2015wga}
%\bibitem{Kharuk:2015wga} 
 I.~Kharuk and S.~M.~Sibiryakov,
  %``Emergent Lorentz invariance with chiral fermions,''
  Theor.\ Math.\ Phys.\  {\bf 189}, no. 3, 1755 (2016)
  [Teor.\ Mat.\ Fiz.\  {\bf 189}, no. 3, 405 (2016)]
%  doi:10.1134/S0040577916120084
  [arXiv:1505.04130 [hep-th]].
  %%CITATION = ARXIV:1505.04130;%%
  %1 citations counted in INSPIRE as of 14 Nov 2015

%\cite{Janiszewski:2012nb}
\bibitem{Janiszewski:2012nb} 
  S.~Janiszewski and A.~Karch,
  %``Non-relativistic holography from Horava gravity,''
  JHEP {\bf 1302}, 123 (2013)
  [arXiv:1211.0005 [hep-th]];
  %%CITATION = ARXIV:1211.0005;%%
  %32 citations counted in INSPIRE as of 14 Nov 2015
%\cite{Janiszewski:2012nf}
%\bibitem{Janiszewski:2012nf} 
%  S.~Janiszewski and A.~Karch,
  %``String Theory Embeddings of Nonrelativistic Field Theories and
  %Their Holographic Ho\v rava Gravity Duals,''
  Phys.\ Rev.\ Lett.\  {\bf 110}, no. 8, 081601 (2013)
  [arXiv:1211.0010 [hep-th]].
  %%CITATION = ARXIV:1211.0010;%%
  %21 citations counted in INSPIRE as of 14 Nov 2015

%\cite{Griffin:2012qx}
\bibitem{Griffin:2012qx} 
  T.~Griffin, P.~Ho\v rava and C.~M.~Melby-Thompson,
  %``Lifshitz Gravity for Lifshitz Holography,''
  Phys.\ Rev.\ Lett.\  {\bf 110}, no. 8, 081602 (2013)
  [arXiv:1211.4872 [hep-th]].
  %%CITATION = ARXIV:1211.4872;%%
  %48 citations counted in INSPIRE as of 14 Nov 2015

%\cite{Anselmi:2007ri}
\bibitem{Anselmi:2007ri} 
  D.~Anselmi and M.~Halat,
  %``Renormalization of Lorentz violating theories,''
  Phys.\ Rev.\ D {\bf 76}, 125011 (2007)
  [arXiv:0707.2480 [hep-th]].
  %%CITATION = ARXIV:0707.2480;%%
  %86 citations counted in INSPIRE as of 14 Nov 2015

%\cite{Fujimori:2015mea}
\bibitem{Fujimori:2015mea} 
  T.~Fujimori, T.~Inami, K.~Izumi and T.~Kitamura,
  %``Tree-Level Unitarity and Renormalizability in Lifshitz Scalar Theory,''
  PTEP {\bf 2016}, no. 1, 013B08 (2016)
%  doi:10.1093/ptep/ptv185
  [arXiv:1510.07237 [hep-th]].
  %%CITATION = ARXIV:1510.07237;%%

%\cite{Anselmi:2008bq}
\bibitem{Anselmi:2008bq} 
  D.~Anselmi,
  %``Weighted power counting and Lorentz violating gauge theories. I. General properties,''
  Annals Phys.\  {\bf 324}, 874 (2009)
  [arXiv:0808.3470 [hep-th]].
  %%CITATION = ARXIV:0808.3470;%%
  %71 citations counted in INSPIRE as of 14 Nov 2015

%\cite{Blas:2011ni}
\bibitem{Blas:2011ni} 
  D.~Blas and S.~Sibiryakov,
  %``Horava gravity versus thermodynamics: The Black hole case,''
  Phys.\ Rev.\ D {\bf 84}, 124043 (2011)
  [arXiv:1110.2195 [hep-th]].
  %%CITATION = ARXIV:1110.2195;%%
  %46 citations counted in INSPIRE as of 14 Nov 2015

%\cite{Orlando:2009en}
\bibitem{Orlando:2009en} 
  D.~Orlando and S.~Reffert,
  %``On the Renormalizability of Horava-Lifshitz-type Gravities,''
  Class.\ Quant.\ Grav.\  {\bf 26}, 155021 (2009)
  [arXiv:0905.0301 [hep-th]].
  %%CITATION = ARXIV:0905.0301;%%
  %123 citations counted in INSPIRE as of 15 Nov 2015

%\cite{Horava:2009if}
\bibitem{Horava:2009if} 
  P.~Horava,
  %``Spectral Dimension of the Universe in Quantum Gravity at a
  %Lifshitz Point,'' 
  Phys.\ Rev.\ Lett.\  {\bf 102}, 161301 (2009)
  [arXiv:0902.3657 [hep-th]]; 
  %%CITATION = ARXIV:0902.3657;%%
  %413 citations counted in INSPIRE as of 15 Nov 2015
%\cite{Anderson:2011bj}
%\bibitem{Anderson:2011bj} 
  C.~Anderson, S.~J.~Carlip, J.~H.~Cooperman, P.~Horava, R.~K.~Kommu and P.~R.~Zulkowski,
  %``Quantizing Horava-Lifshitz Gravity via Causal Dynamical Triangulations,''
  Phys.\ Rev.\ D {\bf 85}, 044027 (2012)
  [arXiv:1111.6634 [hep-th]].
  %%CITATION = ARXIV:1111.6634;%%
  %49 citations counted in INSPIRE as of 15 Nov 2015

%\cite{Ambjorn:2010hu}
\bibitem{Ambjorn:2010hu} 
  J.~Ambjorn, A.~Gorlich, S.~Jordan, J.~Jurkiewicz and R.~Loll,
  %``CDT meets Horava-Lifshitz gravity,''
  Phys.\ Lett.\ B {\bf 690}, 413 (2010)
  [arXiv:1002.3298 [hep-th]].
  %%CITATION = ARXIV:1002.3298;%%
  %62 citations counted in INSPIRE as of 15 Nov 2015

%\cite{Sotiriou:2011mu}
\bibitem{Sotiriou:2011mu} 
  T.~P.~Sotiriou, M.~Visser and S.~Weinfurtner,
  %``Spectral dimension as a probe of the ultraviolet continuum regime of causal dynamical triangulations,''
  Phys.\ Rev.\ Lett.\  {\bf 107}, 131303 (2011)
  [arXiv:1105.5646 [gr-qc]].
  %%CITATION = ARXIV:1105.5646;%%
  %42 citations counted in INSPIRE as of 15 Nov 2015

%\cite{Benedetti:2014dra}
\bibitem{Benedetti:2014dra} 
  D.~Benedetti and J.~Henson,
  %``Spacetime condensation in (2+1)-dimensional CDT from a Ho\v rava-Lifshitz minisuperspace model,''
  Class.\ Quant.\ Grav.\  {\bf 32}, no. 21, 215007 (2015)
  [arXiv:1410.0845 [gr-qc]].
  %%CITATION = ARXIV:1410.0845;%%
  %2 citations counted in INSPIRE as of 15 Nov 2015

%\cite{Benedetti:2013pya}
\bibitem{Benedetti:2013pya} 
  D.~Benedetti and F.~Guarnieri,
  %``One-loop renormalization in a toy model of Ho\v rava-Lifshitz gravity,''
  JHEP {\bf 1403}, 078 (2014)
  [arXiv:1311.6253 [hep-th]].
  %%CITATION = ARXIV:1311.6253;%%
  %9 citations counted in INSPIRE as of 15 Nov 2015

%\cite{D'Odorico:2014iha}
\bibitem{D'Odorico:2014iha} 
  G.~D'Odorico, F.~Saueressig and M.~Schutten,
  %``Asymptotic Freedom in Ho\v rava-Lifshitz Gravity,''
  Phys.\ Rev.\ Lett.\  {\bf 113}, no. 17, 171101 (2014)
  [arXiv:1406.4366 [gr-qc]];
  %%CITATION = ARXIV:1406.4366;%%
  %9 citations counted in INSPIRE as of 15 Nov 2015
%\cite{D'Odorico:2015yaa}
%\bibitem{D'Odorico:2015yaa} 
  G.~D'Odorico, J.~W.~Goossens and F.~Saueressig,
  %``Covariant computation of effective actions in Ho\v rava-Lifshitz gravity,''
  JHEP {\bf 1510}, 126 (2015)
  [arXiv:1508.00590 [hep-th]].
  %%CITATION = ARXIV:1508.00590;%%

%\cite{Giribet:2010th}
\bibitem{Giribet:2010th} 
  G.~Giribet, D.~L.~Nacir and F.~D.~Mazzitelli,
  %``Counterterms in semiclassical Horava-Lifshitz gravity,''
  JHEP {\bf 1009}, 009 (2010)
  %doi:10.1007/JHEP09(2010)009
  [arXiv:1006.2870 [hep-th]].
  %%CITATION = doi:10.1007/JHEP09(2010)009;%%
  %13 citations counted in INSPIRE as of 04 Dec 2015

%\cite{Nesterov:2010yi}
\bibitem{Nesterov:2010yi} 
  D.~Nesterov and S.~N.~Solodukhin,
  %``Gravitational effective action and entanglement entropy in UV modified theories with and without Lorentz symmetry,''
  Nucl.\ Phys.\ B {\bf 842}, 141 (2011)
  %doi:10.1016/j.nuclphysb.2010.08.006
  [arXiv:1007.1246 [hep-th]].
  %%CITATION = doi:10.1016/j.nuclphysb.2010.08.006;%%
  %17 citations counted in INSPIRE as of 04 Dec 2015

%\cite{Baggio:2011ha}
\bibitem{Baggio:2011ha} 
  M.~Baggio, J.~de Boer and K.~Holsheimer,
  %``Anomalous Breaking of Anisotropic Scaling Symmetry in the Quantum Lifshitz Model,''
  JHEP {\bf 1207}, 099 (2012)
  %doi:10.1007/JHEP07(2012)099
  [arXiv:1112.6416 [hep-th]].
  %%CITATION = doi:10.1007/JHEP07(2012)099;%%
  %31 citations counted in INSPIRE as of 04 Dec 2015

%\cite{Sotiriou:2009gy}
\bibitem{Sotiriou:2009gy} 
  T.~P.~Sotiriou, M.~Visser and S.~Weinfurtner,
  %``Phenomenologically viable Lorentz-violating quantum gravity,''
  Phys.\ Rev.\ Lett.\  {\bf 102}, 251601 (2009)
%  doi:10.1103/PhysRevLett.102.251601
  [arXiv:0904.4464 [hep-th]].
  %%CITATION = doi:10.1103/PhysRevLett.102.251601;%%
  %202 citations counted in INSPIRE as of 17 Nov 2015

%\cite{Blas:2009yd}
\bibitem{Blas:2009yd} 
  D.~Blas, O.~Pujolas and S.~Sibiryakov,
  %``On the Extra Mode and Inconsistency of Horava Gravity,''
  JHEP {\bf 0910}, 029 (2009)
% doi:10.1088/1126-6708/2009/10/029
  [arXiv:0906.3046 [hep-th]].
  %%CITATION = doi:10.1088/1126-6708/2009/10/029;%%
  %248 citations counted in INSPIRE as of 17 Nov 2015

%\cite{Becchi:1974md}
\bibitem{Becchi:1974md} 
  C.~Becchi, A.~Rouet and R.~Stora,
  %``Renormalization of the Abelian Higgs-Kibble Model,''
  Commun.\ Math.\ Phys.\  {\bf 42}, 127 (1975); 
%  doi:10.1007/BF01614158
  %%CITATION = doi:10.1007/BF01614158;%%
  %934 citations counted in INSPIRE as of 04 Dec 2015
%\cite{Becchi:1975nq}
%\bibitem{Becchi:1975nq} 
%  C.~Becchi, A.~Rouet and R.~Stora,
  %``Renormalization of Gauge Theories,''
  Annals Phys.\  {\bf 98}, 287 (1976).
%  doi:10.1016/0003-4916(76)90156-1
  %%CITATION = doi:10.1016/0003-4916(76)90156-1;%%
  %1315 citations counted in INSPIRE as of 04 Dec 2015

\bibitem{Tyutin}
I.V.~Tyutin, ``Gauge Invariance in Field Theory and Statistical
Physics in Operator Formalism,'' Lebedev Institute preprint N39 (1975)
[arXiv:0812.0580 [hep-th]].


\bibitem{Weinberg}
S.~Weinberg, ``The Quantum Theory of Fields. Volume II: Modern
Applications,'' Cambridge University Press 1996. 

\bibitem{DeWitt}
B.S.~DeWitt, ``Dynamical Theory of Groups and Fields,'' Gordon and
Breach, 1965.

%\cite{Veltman:1975vx}
\bibitem{Veltman:1975vx} 
  M.~J.~G.~Veltman,
  ``Quantum Theory of Gravitation,''
%in ``Les Houches 1975, Proceedings, Methods In Field Theory,''
% Amsterdam 1976, 265-327.
  Conf.\ Proc.\ C {\bf 7507281}, 265 (1975).
  %%CITATION = CONFP,C7507281,265;%%
  %12 citations counted in INSPIRE as of 04 Dec 2015

%\cite{Abbott:1981ke}
\bibitem{Abbott:1981ke} 
  L.~F.~Abbott,
  %``Introduction to the Background Field Method,''
  Acta Phys.\ Polon.\ B {\bf 13}, 33 (1982).
  %%CITATION = APPOA,B13,33;%%
  %253 citations counted in INSPIRE as of 04 Dec 2015

%\cite{Barvinsky:2017zlx}
\bibitem{Barvinsky:2017zlx} 
  A.~O.~Barvinsky, D.~Blas, M.~Herrero-Valea, S.~M.~Sibiryakov and
  C.~F.~Steinwachs, 
  ``Renormalization of gauge theories in the background-field approach,''
  arXiv:1705.03480 [hep-th].
  %%CITATION = ARXIV:1705.03480;%%
  %2 citations counted in INSPIRE as of 23 Jun 2017

\bibitem{Abbott:1980hw} 
  L.~F.~Abbott,
 % ``The Background Field Method Beyond One Loop,''
Nucl.\ Phys.\ B {\bf 185}, 189 (1981).
%doi:10.1016/0550-3213(81)90371-0

%\cite{Barvinsky:1988ds}
\bibitem{Barvinsky:1988ds} 
  A.~O.~Barvinsky and G.~A.~Vilkovisky,
  ``The Effective Action In Quantum Field Theory: Two Loop Approximation,''
  IN *BATALIN, I.A. (ED.) ET AL.: QUANTUM FIELD THEORY AND QUANTUM
  STATISTICS, VOL. 1*, 245-275 (1988).

\bibitem{Grassi:1995wr} 
  P.~A.~Grassi,
 % ``Stability and renormalization of Yang-Mills theory with
 % background field method: A Regularization independent proof,'' 
Nucl.\ Phys.\ B {\bf 462}, 524 (1996)
% doi:10.1016/0550-3213(96)00017-X
[hep-th/9505101].

\bibitem{Anselmi:2013kba} 
  D.~Anselmi,
%  ``Background field method, Batalin-Vilkovisky formalism and
%  parametric completeness of renormalization,'' 
Phys.\ Rev.\ D {\bf 89}, no. 4, 045004 (2014)
% doi:10.1103/PhysRevD.89.045004
[arXiv:1311.2704 [hep-th]].

%\cite{Sotiriou:2011dr}
\bibitem{Sotiriou:2011dr} 
  T.~P.~Sotiriou, M.~Visser and S.~Weinfurtner,
  %``Lower-dimensional Horava-Lifshitz gravity,''
  Phys.\ Rev.\ D {\bf 83}, 124021 (2011)
 % doi:10.1103/PhysRevD.83.124021
  [arXiv:1103.3013 [hep-th]].
  %%CITATION = doi:10.1103/PhysRevD.83.124021;%%
  %13 citations counted in INSPIRE as of 29 Nov 2015

\end{thebibliography}
\end{document}